\documentclass[letterpaper, onecolumn,draftcls]{IEEEtran}
\usepackage{graphics,color,subfigure,latexsym,amssymb,amsmath,amsthm,url}
\usepackage{bbm}
\usepackage[pdftex]{graphicx}
\newcommand{\eqdef} {\mathrel{\mathop:}=}
\newtheorem{definition}{Definition}
\newtheorem{remark}{Remark}
\newtheorem{theorem}{Theorem}

\newtheorem{lemma}{Lemma}
\newtheorem{proposition}{Proposition}

\newcommand{\ben}{\begin{enumerate}}
\newcommand{\een}{\end{enumerate}}
\newcommand{\bea}{\begin{eqnarray}}
\newcommand{\eea}{\end{eqnarray}}
\newcommand{\bean}{\begin{eqnarray*}}
\newcommand{\eean}{\end{eqnarray*}}

\newcommand{\cX}{{\cal X}}

\newcommand{\cY}{{\cal Y}}

\newcommand{\cA}{{\cal A}}

\newcommand{\cP}{{\cal P}}
\newcommand{\cB}{{\cal B}}
\newcommand{\cS}{{\cal S}}

\newcommand{\cM}{{\cal M}}
\newcommand{\cG}{\cite[p.~94]{gallager68}}

\allowdisplaybreaks

\def\b0{\mathbf{0}}
\def\bX{\mathbf{X}}

\def\bY{\mathbf{Y}}

\def\bbR{\mathbb{R}}
\def\bbZ{\mathbb{Z}}
\def\bx{\mathbf{x}}
\def\by{\mathbf{y}}
\def\bz{\mathbf{z}}

\def\bb{\mathbf{b}}

\def\mE{\textnormal{E}}
\def\mP{\textnormal{P}}
\def\me{\textnormal{e}}
\def\mcr{\textnormal{cr}}
\def\mo{\textnormal{o}}
\def\mr{\textnormal{r}}
\def\mI{I}
\def\mD{\textnormal{D}}
\def\mV{\textnormal{Var}}
\def\mSP{\textnormal{SP}}
\def\supp{\textnormal{supp}}
\def\b1{\mathbbm{1}}
\def\W1{W_{1^{-},P,R}}

\begin{document}
\title{On exact asymptotics of the error probability \\ in channel coding: symmetric channels}
\author{Y\"{u}cel Altu\u{g}~\IEEEmembership{}and Aaron B.~Wagner,~\IEEEmembership{Senior Member, IEEE}
\thanks{The material in this paper was presented in part at the 49th Annual Allerton Conference on Communications, Control, and Computing, 2013 Information Theory
and Applications Workshop, and 2014 IEEE International Symposium on Information Theory.

Y\"{u}cel Altu\u{g} was, and Aaron B. Wagner is, with the School of Electrical and Computer Engineering, Cornell University, Ithaca, NY 14853. (E-mail: {\em ya68@cornell.edu, wagner@ece.cornell.edu}).}}
\maketitle
\begin{abstract}
The exact order of the optimal sub-exponentially decaying factor in the classical bounds on the error probability of fixed-length codes over a Gallager-symmetric discrete memoryless channel with and without ideal feedback is determined. Regardless of the availability of feedback, it is shown that the order of the optimal sub-exponential factor exhibits a dichotomy. Moreover, the proof technique is used to establish the third-order term in the normal approximation for symmetric channels, where a similar dichotomy is shown to exist.
\end{abstract}

\section{Introduction}
\label{sec:intro}

In channel coding, error exponents describe the rate of decay of the error probability with the rate held fixed below the capacity (e.g., \cite{feinstein54}--\cite{csiszar-korner81} and references therein). As such, they provide an exponentially fast convergence result in the channel coding theorem, and thereby indicate approximately how large of a blocklength one needs to achieve a target error probability for a given rate. The caveat with classical error exponent results, however, is that they are typically expressed as bounds on the reliability function, which is defined as (e.g., \cite[Eq.~(5.8.8)]{gallager65})
\begin{equation}
\mE(R) \eqdef \limsup_{N \to \infty}-\frac{1}{N}\ln \mP_{\me}(N,R),	
\end{equation}
where $\mP_\me(N,R)$ is the minimum error probability of all codes with blocklength $N$ and rate $R$. Thus, they ignore the sub-exponential factors in $\mP_{\me}(N,R)$, which potentially could be quite significant for small to moderate $N$. This is especially true for rates near capacity, since typically both the exponent and its first derivative vanish as the rate approaches capacity. Therefore, one would like to have more refined bounds on $\mP_{\me}(N,R)$ that capture the sub-exponential factors, which we will also refer to as the \emph{pre-factor(s)}.

Classical bounds on the pre-factor were quite loose. In particular, until recently the best known upper and lower bounds on the optimal pre-factor that are valid for any DMC were $\textnormal{O}(1)$ and $\Omega(N^{-|\cX||\cY|})$, due to Fano~\cite{fano61} and Haroutunian~\cite{haroutunian68}, respectively. Here, $|\cX|$ and $|\cY|$ denote the cardinality of the input and output alphabet of the channel, respectively. The authors have improved upon these results to obtain relatively tight bounds on the order of the pre-factor, which we summarize next. Specifically, \cite{altug12a} proves that the error probability of any $(N,R)$ constant composition code, i.e., a code in which all codewords possess the same empirical distribution, is lower bounded by
\begin{equation}
\frac{K_1}{N^{\frac{1}{2}(1+|\mE_{\mSP}^{\prime}(R)|)}}\textnormal{e}^{-N \mE_{\mSP}(R)},
\label{eq:SP-asym}	
\end{equation}
where $\mE_{\mSP}^\prime(R)$ is the slope of the sphere-packing exponent at $R$ and $K_1 \in \bbR^+$ is a constant that depends on the channel and $R$. In \cite{altug13-a}, it is shown that if the channel satisfies a certain condition, then the optimal error probability is upper bounded by
\begin{equation}
\frac{K_2}{N^{\frac{1}{2}(1+\bar{\rho}_R)}}\textnormal{e}^{-N \mE_{\textnormal{r}}(R)},	
\end{equation}
where $\bar{\rho}_R$ is related to the slope of the random coding exponent and is typically equal to $|\mE^{\prime}_{\textnormal{r}}(R)|$, and $K_2 \in \bbR^+$ is a constant that depends on the channel and $R$. For the remaining small class of channels, the following upper bound holds
\begin{equation}
\frac{K_3}{\sqrt{N}}\textnormal{e}^{-N \mE_{\textnormal{r}}(R)},	
\end{equation}
where $K_3 \in \bbR^+$ is a constant that depends on the channel and $R$.
Note that the order of the aforementioned upper and lower bounds asymptotically coincide as the rate approaches capacity.

Related to the above bounds, one of the classical results of Elias is worth mentioning. In~\cite{elias55}, he considered binary symmetric and erasure channels and proved that the order of the optimal pre-factor for the binary symmetric (resp. erasure) channel is $\Theta(N^{-\frac{1}{2}(1+|\textnormal{E}^\prime(R)|)})$ (resp. $\Theta(N^{-\frac{1}{2}})$) for rates above the critical rate, where $\textnormal{E}^\prime(R)$ is the slope of the reliability function.

In this paper, we show that for the class of symmetric channels (see Definition~\ref{def:symmetric} to follow) we can improve the bounds in \cite{altug12a} and \cite{altug13-a} to give an exact characterization of the order of the dominant sub-exponential factor. Specifically, we prove a dichotomy of symmetric channels in terms of the order of their optimal pre-factors. For the typical symmetric channels, which we call \emph{nonsingular channels}, the optimal order is $\Theta(N^{-\frac{1}{2}(1+|\mE^\prime(R)|)})$, whereas for the remaining symmetric channels, namely \emph{singular channels}, $\Theta(N^{-\frac{1}{2}})$ is the optimal order. These results imply that every symmetric channel has a pre-factor order that matches either that of the BEC or that of the BSC. Thus, Elias had already found all of the different orders that can occur for symmetric channels.

For both singular and nonsingular channels, the upper bound on
the pre-factor follows from~\cite{altug13-a} (which has been 
strengthened in several ways~\cite{Honda:RC:ISIT15,Honda:RC:ISIT18,Scarlett:RC:Allerton13,Scarlett:RC:Mismatched,Font:RCU}). 
Our contribution is improving the lower bound on the order of the pre-factor, i.e., obtaining a better pre-factor in the sphere-packing bound. There are multiple ways of proving the sphere-packing bound, some more amenable to obtaining pre-factor bounds than the others. For a comparison of these techniques, see \cite[Section~III.A]{altug12a}. Among these methods, the one that relates the error probability of a given code to the error probability of a related binary hypothesis test with the aid of an auxiliary output distribution is well suited for pre-factor analysis. This method can be traced back to at least the classical results of Blahut~\cite{blahut74} and is the starting point of the derivation of \eqref{eq:SP-asym}. However, the auxiliary output distribution used in \cite{altug12a} does not admit a simple explicit form. Indeed, it is defined by using the saddle-point of a certain optimization problem, which is intimately related to the sphere-packing exponent. This complication is due to the asymmetry of the channel. Once we restrict our attention to symmetric channels, it is possible to show a simple characterization of this distribution (see \eqref{eq:qR} and Proposition~\ref{lem:SP-optimizers-new} to follow), which is in the form of a \emph{tilted distribution}. Since this distribution is independent of the code, we can dispense with the constant composition assumption\footnote{The possibility of proving the sphere-packing bound without the constant composition restriction for symmetric channels was first observed in \cite{wei-sason08}, where the proof methodology of Shannon~\emph{et al.}~\cite{SGB67} was followed.} in \cite{altug12a}.

For the singular case, we introduce a new method of proving the sphere-packing bound. The idea is the following: consider any singular symmetric channel $W$ and any $(N,R)$ code over $W$. Let $\mathcal{E}$ denote the event that the code makes an error. Define the \emph{information density}
\begin{equation}
\imath(x;y) \eqdef \ln \frac{W(y|x)}{\sum_{z \in \cX}\frac{W(y|z)}{|\cX|}}.
\end{equation}
By using Wolfowitz's strong converse (e.g., \cite{wolfowitz57}), one can argue that
\begin{equation}
\Pr\left[ \mathcal{E} \, \bigg| \, \sum_{n=1}^N \imath(X_n;Y_n) \leq R \right] \approx 1,	
\label{eq:intro-1}
\end{equation}
where the probability is induced by the uniform distribution over the messages and the channel, and $\bX^N$ (resp. $\bY^N$) denotes the input (resp. output) of the channel. Hence,
\begin{align}
\Pr[\mathcal{E}] & \geq  \Pr\left[ \sum_{n=1}^N \imath(X_n;Y_n) \leq R \right]\Pr\left[ \mathcal{E} \, \bigg| \, \sum_{n=1}^N \imath(X_n;Y_n) \leq R \right] \\
& \approx \Pr\left[ \sum_{n=1}^N \imath(X_n;Y_n) \leq R \right]. \label{eq:intro-2}
\end{align}
Due to the symmetry of $W$, the random variables in \eqref{eq:intro-2} can be shown to be independent and identically distributed (i.i.d.), and hence one can apply classical exact asymptotics results (e.g., \cite{bahadur-rao60}) to deduce an exponentially decaying lower bound with a pre-factor order of $1/\sqrt{N}$. However, this procedure results in a useful lower bound only if the exponent matches the reliability function, i.e., one needs
\begin{equation}
\lim_{N \to \infty}-\frac{1}{N}\ln \Pr\left[ \sum_{n=1}^N \imath(X_n;Y_n) \leq R \right] = \textnormal{E}_{\textnormal{SP}}(R). \label{eq:intro-3}	
\end{equation}
Although \eqref{eq:intro-3} is not true in general, it can be shown to be so for singular and symmetric channels, thus we can deduce an exponentially vanishing lower bound with the sphere-packing exponent and $\Theta(1/\sqrt{N})$ as the dominant sub-exponential factor.

Furthermore, we show that for both singular and nonsingular symmetric channels the pre-factor order is not affected by the presence of ideal feedback. It is well known that for symmetric channels, feedback does not improve the reliability function above the critical rate (e.g., \cite{haroutunian77}). The results herein strengthen this statement to assert that both the exponent and the dominant sub-exponential factor are unaffected by feedback. For asymmetric channels, see Nakibo\u{g}lu~\cite{Nakiboglu:SPB:IT,Nakiboglu:SPB:TCOM} and Wagner \emph{et al.}~\cite{Altug:ISIT14,Shende:Noisy:Allerton,Shende:Feedback:ISIT,Wagner:TimidBold} for the effect of feedback in the error exponent and normal approximation regimes, respectively.

Moreover, we also apply the aforementioned proof technique to characterize the third-order term in the normal approximation for singular channels. Specifically, for singular and symmetric channels, we prove a converse result, which is valid in the presence of feedback, which implies a dichotomy of the third-order term in the normal approximation for symmetric channels once coupled with \cite{tomamichel-tan13} and \cite[Sec.~3.4.5]{polyanskiy10-phd}. A remarkable aspect of this dichotomy is that its defining property is again singularity of the channel.

We conclude this section by noting that the type of symmetry notion is crucial regarding the dichotomy of the optimal pre-factor of the symmetric channels. Specifically, if one considers \emph{strongly symmetric channels}, i.e., if every row (resp. column) of the channel is a permutation of every other row (resp. column), which is a proper subset of symmetric channels we consider in this paper, then one can show that (e.g., \cite{Dobrushin62}) $\Theta(N^{-\frac{1}{2}(1+|\textnormal{E}^\prime(R)|)})$ is the order of the optimal pre-factor for rates above the critical rate. Evidently, there is no dichotomy for this class of channels, since it is not rich enough to include singular channels (see Remark~\ref{rem:misc}(iii) to follow). Finally, it is possible to extract the constants from our proofs to obtain finite blocklength bounds on the error probability. However, the resulting expressions are rather complicated, so we shall state the results in asymptotic form to elucidate the dichotomy.


\section{Notation, Definitions and Statement of the Results}
\label{sec:notation-definitions}
\subsection{Notation}
\label{ssec:notation}
Boldface letters denote vectors, and regular letters with subscripts denote individual components of vectors. Furthermore, capital letters represent random variables, and lowercase letters denote individual realizations of the corresponding random variable. For a finite set $\cA$, $\cP(\cA)$ (resp. $U_{\cA}$) denotes the set of all probability measures (resp.\ the uniform probability measure) on $\cA$. Similarly, for two finite sets $\cA$ and $\cB$, $\cP(\cB|\cA)$ denotes the set of all stochastic matrices from $\cA$ to $\cB$. Given any $P \in \cP(\cA)$, $\supp(P) \eqdef \{ a \in \cA \, : \, P(a) >0\}$. $\b1{\{ \cdot \} }$ denotes the standard indicator function. Given probability measures $\lambda_{1}$ and $\lambda_{2}$, $\lambda_{1} \ll \lambda_{2}$ means that $\lambda_{1}$ is absolutely continuous with respect to $\lambda_{2}$ (that is, $\lambda_{2}$ dominates  $\lambda_{1}$) and $\lambda_{1} \equiv \lambda_{2}$ means that $\lambda_{1} \ll \lambda_{2} $ and $\lambda_{2} \ll \lambda_{1}$. $\Phi(\cdot)$ (resp. $\phi(\cdot)$) denotes the cumulative distribution function (resp.\ probability density function) of the standard Gaussian random variable. $\bbZ^+$, $\bbR, \bbR^{+}$ and $\bbR_{+}$ denote the set of positive integers, reals, positive reals and non-negative reals, respectively. We follow the notation of the book of Csisz\'ar-K\"orner~\cite{csiszar-korner81} for standard information theoretic quantities.
\subsection{Definitions}
\label{ssec:defn}
An $(N,R)$ \emph{code}, say $(f,\varphi)$, consists of an encoder, i.e., $f \colon \cM \to \cX^N$, where $\cM \eqdef \{ 1, \ldots, \lceil e^{NR} \rceil\}$ is the set of messages to be transmitted, and a decoder, i.e., $\varphi \colon \cY^N \to \cM$. Let $\{\mathcal{A}_m\}_{m=1}^{|\mathcal{M}|}$ denote the decoding regions and $\bar{\mP}_{\me}(f, \varphi)$ denote the average error probability of $(f, \varphi)$. Evidently,
\begin{equation}
\bar{\mP}_{\me}(f, \varphi) = \frac{1}{|\cM|}\sum_{m \in \cM} \sum_{\by^N \in \cA_m^c}P_{\bY^N|\bX^N}(\by^N | f(m)). \label{eq:Pe-nonfeedback}
\end{equation}
$\bar{\mP}_{\me}(N,R)$ denotes the minimum average error probability attainable by any $(N,R)$ code.
Similarly, $\mP_{\me}(N,R)$ denotes the minimum maximal error probability attainable by any $(N,R)$ code.

For any $\epsilon \in (0,1)$,
\begin{align}
M^\ast(N, \epsilon) & \eqdef \max\{\lceil e^{NR} \rceil \in \bbR_+ : \bar{\mP}_{\me}(N,R) \leq \epsilon \}, \label{eq:Mstar}\\
M^\ast_{\textnormal{c}}(N, \epsilon) & \eqdef \max\{\lceil e^{NR} \rceil \in \bbR_+ : \bar{\mP}_{\me, \textnormal{c}}(N,R) \leq \epsilon \}, \label{eq:Mstar-c}
\end{align}
where $\bar{\mP}_{\me, \textnormal{c}}(N,R)$ denotes the minimum average error probability attainable by any $(N,R)$ constant composition code.

An $(N,R)$ \emph{code with ideal feedback}, say $(f,\varphi)$, consists of an encoder, i.e., $\{ f_n \colon \cM \times \cY^{n-1} \to \cX \}_{n=1}^N$, where $\cM \eqdef \{ 1, \ldots, \lceil e^{NR} \rceil\}$ is the set of messages to be transmitted, and a decoder, i.e., $\varphi \colon \cY^N \to \cM$. Let $\{\mathcal{A}_m\}_{m=1}^{|\mathcal{M}|}$ denote the decoding regions and $\bar{\mP}_{\me}(f, \varphi)$ denote the average error probability of $(f, \varphi)$. Define
\begin{equation}
P_{\bY^N|M}(\by^N | m) \eqdef \prod_{n=1}^{N}W(y_n|f_{n}(m, \by^{n-1})), \label{eq:HT-W}
\end{equation}
where $f_{n}(m,\by^{n-1})$ denotes the output of the encoder at time $n$ if message $m$ is transmitted, and $\by^{n-1}$ denotes the previous channel outputs, with the usual convention $\by^0 \eqdef \emptyset$. Again, 
\begin{equation}
\bar{\mP}_{\me}(f, \varphi)  = \frac{1}{|\cM|}\sum_{m \in \cM} \sum_{\by^N \in \cA_m^c}P_{\bY^N|M}(\by^N | m). \label{eq:Pe-nonsingular}
\end{equation}
$\bar{\mP}_{\me, \textnormal{fb}}(N,R)$ denotes the minimum average error probability attainable by any $(N,R)$ code with ideal feedback.

Given any channel $W \in \cP(\cY|\cX)$ and $R \in \bbR_+$, we recall the following classical quantities (e.g., \cite[Sec.~2.5]{csiszar-korner81})
\begin{align}
 \mE_{\mSP}(R,Q) & \eqdef \min_{V \in \cP(\cY|\cX) \colon \mI(Q;V) \leq R} \mD(V\|W|Q), \label{eq:SP-H-0}\\
\mE_{\mSP}(R) & \eqdef \max_{Q \in \cP(\cX)}\mE_{\mSP}(R,Q), \label{eq:SP-H-1}\\
 \tilde{\mE}_{\mSP}(R,Q) & \eqdef \sup_{\rho \geq 0} \left\{\mE_{\mo}(\rho, Q)  -\rho R \right\}, \label{eq:SP-SGB-0}\\
 \tilde{\mE}_{\mSP}(R) & \eqdef \max_{Q \in \cP(\cX)}\tilde{\mE}_{\mSP}(R,Q), \label{eq:SP-SGB-1}\\
\mE_{\mr}(R,Q) & \eqdef \max_{0 \leq \rho \leq 1} \left\{ \mE_{\mo}(\rho, Q) -\rho R \right\}, \label{eq:Er-0}\\
\mE_{\mr}(R) & \eqdef \max_{Q \in \cP(\cX)} \mE_{\mr}(R,Q), \label{eq:Er-1}
\end{align}
where
\begin{equation}
\mE_{\mo}(\rho, Q) \eqdef -\ln \sum_{y \in \cY}\left( \sum_{x \in \cX} Q(x)W(y|x)^{1/(1+\rho)}\right)^{1+\rho}. \label{eq:Eo}
\end{equation}

It is well known that given any $R \in \bbR_+$, $\mE_{\mSP}(R,Q) \geq \tilde{\mE}_{\mSP}(R, Q)$ for all $Q \in \cP(\cX)$ and $\mE_{\mSP}(R) = \tilde{\mE}_{\mSP}(R)$ (e.g., \cite[Ex.~2.5.23]{csiszar-korner81}). $R_\infty$ denotes the maximum rate such that for all rates below it, $\mE_{\mSP}(R) = \infty$ (e.g., \cite[pg.~158]{gallager68}). Also, $R_{\mcr}$ denotes the \emph{critical rate} of the channel, i.e., the value such that $\mE_{\mr}(R) = \mE_{\mSP}(R)$ if and only if $R \geq R_{\mcr}$ (e.g., \cite[pg.~160]{gallager68}). Evidently, $\mE_{\mr}(R) = \mE_{\mSP}(R)  = \tilde{\mE}_{\mSP}(R)$ for all $R \geq R_{\mcr}$.

Given $W \in \cP(\cY|\cX)$, $C(W)$ denotes the capacity of the channel. For any $P \in \cP(\cX)$, define 
\begin{equation}
 q_P(y) \eqdef \sum_{x \in \cX}P(x)W(y|x). 
\end{equation}
For notational convenience, let $q$ denote $q_{U_{\cX}}$. Given any $W \in \cP(\cY|\cX)$, $P \in \cP(\cX)$ and $\epsilon \in (0,1)$, define (e.g., \cite[Sec.~3.4]{polyanskiy10-phd})
\begin{align}
V(P,W) & \eqdef \sum_{x,y}P(x)W(y|x) \left[  \ln \frac{W(y|x)}{q_P(y)} -  \sum_{b}W(b|x)\ln\frac{W(b|x)}{q_P(b)}\right]^2, \label{eq:dispersion}\\
V_\epsilon(W) & \eqdef
\begin{cases}
\min_{Q \colon \mI(Q;W) = C(W)}V(Q,W), & \epsilon \in (0,1/2), \\
\max_{Q \colon \mI(Q;W) = C(W)}V(Q,W), & \epsilon \in [1/2,1).
\end{cases} \label{eq:eps-dispersioin}
\end{align}
We call $V_\epsilon(W)$ the \emph{$\epsilon$-dispersion} of the channel
$W$. The \emph{dispersion} refers to $V_\epsilon(W)$ for $\epsilon < 1/2$.

The following definition is the type of symmetry we use in this work.
\begin{definition}[Gallager~\cG] A discrete channel is \emph{symmetric} if the channel outputs can be partitioned into subsets such that within each subset, the matrix of transition probabilities satisfies the following: each row (resp. column) is a permutation of each other row (resp. column).
\label{def:symmetric}
\end{definition}
We delineate symmetric channels with respect to the order of their optimal pre-factors by using the following notion.
\begin{definition}[Singularity]
A symmetric channel $W \in \cP(\cY|\cX)$ is \emph{singular} if
\begin{equation}
\forall \,  (x,y,z) \in \cX \times \cY \times \cX \textnormal{ s.t. } W(y|x)W(y|z)>0, \ \text{we have} \  W(y|x)=W(y|z).
\label{eq:singularity}
\end{equation}
Otherwise, it is called \emph{nonsingular}.
\label{def:singularity}
\end{definition}

For general channels, the definition of singularity is more
involved~\cite[Definition~1]{altug13-a}. That definition reduces to
Definition~\ref{def:singularity} for symmetric channels, however. 
More precisely,
if a symmetric channel is singular according to 
Definition~\ref{def:singularity},
then it is singular at all rates according to~\cite[Definition~1]{altug13-a},
and, if it is nonsingular according to Definition~\ref{def:singularity},
then it is nonsingular at all rates according to~\cite[Definition~1]{altug13-a}.

An equivalent definition of singularity can be given in terms of the following quantity, which is defined in \cite[Sec.~3.4]{polyanskiy10-phd},
\begin{equation}
V^r(P,W) \eqdef \sum_{x,y}P(x)W(y|x) \left[  \ln \frac{W(y|x)}{q_P(y)} - \sum_{z}\frac{P(z)W(y|z)}{q_P(y)}\ln\frac{W(y|z)}{q_P(y)}\right]^2. \label{eq:reverse-dispersion}
\end{equation}
Specifically, for a symmetric channel $W$ and $P \in \cP(\cX)$ with $P(x) >0$ for all $x \in \cX$, $V^r(P, W)=0$ if and only if $W$ is singular. To see this, note that if $P$ has full support then
\begin{equation}
\left[ V^r(P,W) = 0 \right] \Longleftrightarrow \left[ \ln W(y|x) = \sum_{z}\frac{P(z)W(y|z)}{q_P(y)}\ln W(y|z), \, \forall x \in \cX \ \text{and} \ y \in \cY \, \textnormal{ such that } W(y|x)>0 \right]. \label{eq:prelim-pf1}
\end{equation}
In light of Definition~\ref{def:singularity}, the right side of \eqref{eq:prelim-pf1} is equivalent to saying that $W$ is singular.

In \cite[Lemma~52]{polyanskiy10-phd}, it is claimed that
\begin{equation}
\left[ V^r(P,W) = 0 \right] \Longleftrightarrow \left[ \forall \, (x,y,y^\prime) \colon W(y|x)=W(y^\prime|x) \textnormal{ or } P(x)W(y|x)=0 \right].
\label{eq:rem-yury}
\end{equation}
By choosing $P = U_\cX$ and $W$ to be a BEC with parameter $\delta \in (0,1)$, one can verify that $V^r(P,W) = 0$ by elementary calculation. Evidently, this $(P,W)$ pair does not satisfy the right side of \eqref{eq:rem-yury} and hence \eqref{eq:rem-yury} is incorrect.
For more on singularity, see \cite[Remark~1]{altug13-a}.

\subsection{Statement of the results}
\label{ssec:result}
\begin{theorem}
Let $W$ be a symmetric and nonsingular channel with $R_\mcr < C(W)$.
\begin{itemize}
\item[(i)] For any $R_{\mcr} < R < C(W)$ and any $N$,
\begin{equation}
\mP_{\me}(N,R) \leq \frac{K_1}{N^{\frac{1}{2}(1+|\mE_{\mr}^\prime(R)|)}}
       \exp\left\{-N \mE_{\mr}(R)\right\},
\label{eq:nonsingular-ach}
\end{equation}
where $K_1$ is a positive constant that depends on $W$ and $R$.
\item[(ii)] For any $R_{\infty} < R < C(W)$ and any $N$,
\begin{equation}
\bar{\mP}_{\textnormal{e}, \textnormal{fb}}(N,R) \geq \frac{\tilde{K}_1}{N^{\frac{1}{2}(1+|\mE_{\mSP}^\prime(R)|)}}\exp\left\{-N \mE_{\mSP}(R)\right\},
\label{eq:nonsingular-conv}
\end{equation}
where $\tilde{K}_1$ is a positive constant that depends on $W$ and $R$.
\end{itemize}
\label{thrm:nonsingular}
\end{theorem}
\begin{IEEEproof}
Theorem~\ref{thrm:nonsingular} is proven in Section~\ref{ssec:nonsingular}.\IEEEQEDoff
\end{IEEEproof}

\begin{theorem}
Let $W$ be a symmetric and singular channel with $R_\mcr < C(W)$.
\begin{itemize}
\item[(i)] For any $R_{\mcr} < R < C(W)$ and any $N$,
\begin{equation}
\mP_{\me}(N,R) \leq \frac{K_2}{\sqrt{N}}\exp\left\{-N \mE_{\mr}(R)\right\},
\label{eq:singular-ach}
\end{equation}
where $K_2$ is a positive constant that depends on $W$ and $R$.
\item[(ii)] For any $R_{\infty} < R < C(W)$ and any $N$,
\begin{equation}
\bar{\mP}_{\me, \textnormal{fb}}(N,R) \geq \frac{\tilde{K}_2}{\sqrt{N}}\exp\left\{-N \mE_{\mSP}(R)\right\},
\label{eq:singular-conv}
\end{equation}
where $\tilde{K}_2$ is a positive constant that depends on $W$ and $R$.
\end{itemize}
\label{thrm:singular}
\end{theorem}
\begin{IEEEproof}
Theorem~\ref{thrm:singular} is proven in Section~\ref{ssec:singular}.\IEEEQEDoff
\end{IEEEproof}

\begin{remark}
\begin{itemize}
\item[(i)] For any $W \in \cP(\cY|\cX)$, the following three statements are equivalent (e.g.,~\cite[pg.~160]{gallager68}): $R_{\mcr}<C$, $R_\infty  < C$, and the dispersion of $W$ is positive.

\item[(ii)] Recall that at rates above the critical rate, 
$\mE_{\mSP}(R) = \mE_{\mr}(R)$ by definition. Thus the exponents in
(\ref{eq:nonsingular-ach})--(\ref{eq:singular-conv}) are all the same
in this regime.

\item[(iii)] As mentioned in Section~\ref{sec:intro}, if every row (resp. column) of the channel is a permutation of every other row (resp. column), then we call it a \emph{strongly symmetric channel}. When particularized to this class of channels without feedback, Theorem~\ref{thrm:nonsingular} reduces to a result of Dobrushin~\cite{Dobrushin62} by noting the fact that any strongly symmetric channel with $R_{\mcr} < C$ is necessarily nonsingular (e.g., \cite[Footnote~3]{altug13-a}).

\item[(iv)] For rates above the critical rate, the ratios of the upper and lower bounds in Theorems~\ref{thrm:nonsingular} and \ref{thrm:singular} are bounded away from $0$ and $\infty$ as $N \to \infty$. Indeed, we can explicitly deduce the constants in both theorems from their proofs, although they are not optimized since our goal in this work is to prove an order-optimal pre-factor. Nevertheless, it would be interesting to refine the bounds so that their ratio converges to $1$. A first step in this direction is the work of Scarlett \emph{et al.} \cite{scarlett13}, in which the rate dependence of the pre-factor's constant is investigated for the random coding (i.e., upper) bound. See 
Font-Segura~\emph{et al.}~\cite{Font-Segura:Saddlepoint:CISS18}
for an analogous, though nonrigorous, study of the sphere-packing
bound.
\end{itemize}
\label{rem:misc}
\end{remark}

The technique used to prove part (ii) of Theorem~\ref{thrm:singular} can
also be used to prove the next two results, the first of which
fills a gap in the literature on the normal approximation (see 
Theorem~\ref{theorem:normalapproxgap} to follow).

\begin{theorem}
Given $\epsilon \in (0,1)$ and a singular, symmetric $W$ with $V_\epsilon(W)>0$,
for any $N$,
\begin{equation}
\ln M_{\textnormal{fb}}^\ast(N, \epsilon) \leq N \cdot C(W) + \sqrt{N \cdot V_\epsilon(W)}\Phi^{-1}(\epsilon) + K(\epsilon,W),
\label{eq:thrm-sym}
\end{equation}
where $K(\epsilon, W) \in \bbR^+$ is a constant that depends on $\epsilon$ and $W$.
\label{thrm:sym}
\end{theorem}
\begin{IEEEproof}
Given in Section~\ref{ssec:sym}.\IEEEQEDoff
\end{IEEEproof}

\begin{theorem}
Given a singular and asymmetric $W$,
\begin{itemize}
\item[(i)] If $\epsilon \in (0,1/2)$, then for all $N$,
\begin{equation}
\ln M^\ast_{\textnormal{c}}(N, \epsilon) \leq N\cdot C(W) + \sqrt{N \cdot V_\epsilon(W)}\Phi^{-1}(\epsilon) + \tilde{K}(\epsilon,W),
\label{eq:thrm-asym-1}
\end{equation}
where $\tilde{K}(\epsilon, W) \in \bbR^+$ is a constant that depends on $\epsilon$ and $W$.
\item[(ii)] If $\epsilon \in (1/2,1)$ and $V_\epsilon(W) >0$, then for all $N$,
\begin{equation}
\ln M^\ast_{\textnormal{c}}(N, \epsilon) \leq N\cdot C(W) + \sqrt{N \cdot V_\epsilon(W)}\Phi^{-1}(\epsilon) + \hat{K}(\epsilon,W), \label{eq:thrm-asym-2}
\end{equation}
where $\hat{K}(\epsilon, W) \in \bbR^+$ is a constant that depends on $\epsilon$ and $W$.
\end{itemize}
\label{thrm:asym}
\end{theorem}
\begin{IEEEproof}
Given in Section~\ref{ssec:asym}.\IEEEQEDoff
\end{IEEEproof}

Note that the set of asymmetric and singular channels is not empty. For an example, let $\cX  \eqdef \{ 0,1,2\}$, $\cY \eqdef \{ 0, 1, 2, 3\}$ and consider
\begin{equation}
W(y|x) \eqdef
\begin{cases}
2/3, & (x,y) = (0,0), \\
1/6, & (x,y) \in \{ (0,1), (0, 3), (1,3), (2,1) \}, \\
5/6, & (x,y) \in \{ (1, 2), (2, 2)\}, \\
0, & \textnormal{ else}.
\end{cases}
\end{equation}

Theorem~\ref{thrm:sym} completes the proof of the following assertion:

\begin{theorem}
Given a symmetric $W$ and $\epsilon \in (0,1)$,
\begin{itemize}
\item[(a)] If $W$ is nonsingular and $V_\epsilon(W)>0$, then
\begin{equation}
\ln M^\ast(N,\epsilon) = N\cdot C(W) + \sqrt{N \cdot V_\epsilon(W)}\Phi^{-1}(\epsilon) + \ln\sqrt{N} + \Theta(1).
\label{eq:nonsingular}
\end{equation}
\item[(b)] If $W$ is singular and $V_\epsilon(W) >0$, then
\begin{equation}
\ln M^\ast(N,\epsilon) = N\cdot C(W) + \sqrt{N \cdot V_\epsilon(W)}\Phi^{-1}(\epsilon) + \Theta(1).
\label{eq:singular}
\end{equation}
\item[(c)] If $V_\epsilon(W)=0$, then
\begin{equation}
\ln M^\ast(N,\epsilon) = N\cdot C(W) + \Theta(1).
\label{eq:peculiar}
\end{equation}
\end{itemize}
\label{theorem:normalapproxgap}
\end{theorem}

Specifically, achievability of item (a) follows from \cite[Corollary~54]{polyanskiy10-phd}. The converse of item (a) follows from \cite[Theorem~55]{polyanskiy10-phd}. Achievability of item (b) follows from \cite[Theorem~47]{polyanskiy10-phd}, coupled with Lemma~\ref{lem:sym-prelim}(ii) to follow. The converse for item (b) is proven in Theorem~\ref{thrm:sym}. Item (c) is proven in \cite[Corollary~57]{polyanskiy10-phd}. 

For bounds on the constant in (\ref{eq:nonsingular}), 
              see Moulin~\cite{Moulin:FewNats}.

We assume that the dispersion is positive in Theorem~\ref{thrm:asym}(ii) 
in order to exclude exotic channels; this allows us to focus
on the role of singularity. See \cite[p.~68]{polyanskiy10-phd}
and \cite[Section III]{tomamichel-tan13} for a discussion of
exotic channels.

\section{Proofs}

First, we state two results that are used in the proofs of both Theorems~\ref{thrm:nonsingular} and \ref{thrm:singular}. To this end, for any symmetric channel $W \in \cP(\cY|\cX)$ with $R_{\mcr}<C(W)$ and any $R_{\infty}<R<C(W)$, define
\begin{align}
\bbR^+ \ni \rho_R & \eqdef -\left. \frac{\partial \mE_{\mSP}(r,U_\cX)}{\partial r}\right|_{r=R}, \label{eq:rhoR} \\
\forall \, y \in \cY, q_R(y) & \eqdef \frac{\left(  \sum_{x \in \cX}U_\cX(x)W(y|x)^{\frac{1}{1+\rho_R}}  \right)^{1+\rho_R}}{\sum_{b \in \cY}\left(  \sum_{a \in \cX}U_\cX(a)W(b|a)^{\frac{1}{1+\rho_R}}  \right)^{1+\rho_R}}, \label{eq:qR}
\end{align}
where \eqref{eq:rhoR} is well-defined thanks to \cite[Proposition~3]{altug12a}, and its positivity can be verified by using the fact that $\mE_{\mSP}(R) >0$.
\begin{proposition}
Fix a symmetric channel $W \in \cP(\cY|\cX)$ with $R_{\mcr}<C(W)$. Consider any $R_\infty < R< C(W)$.
\begin{itemize}
\item[(i)]
\begin{equation}
\mE_{\mSP}(R) = \mE_{\mSP}(R,U_\cX) = \tilde{\mE}_{\mSP}(R,U_\cX) = \tilde{\mE}_{\mSP}(R). \label{eq:SP-optimizers-item-i}
\end{equation}
\item[(ii)] For any $\rho \in \bbR_+$,
\begin{equation}
\sum_{y \in \cY}W(y|x)^{\frac{1}{1+\rho}}\left( \sum_{z \in \cX}U_{\cX}(z)W(y|z)^{\frac{1}{1+\rho}}\right)^{\rho} = \sum_{y \in \cY}\left( \sum_{z \in \cX}U_{\cX}(z)W(y|z)^{\frac{1}{1+\rho}}\right)^{1+\rho},
\label{eq:SP-optimizers-1}
\end{equation}
for all $x \in \cX$.
\item[(iii)] $\rho_R$ attains the supremum in the definition of $\tilde{\mE}_{\mSP}(R,U_\cX)$, i.e., \eqref{eq:SP-SGB-1}.
\item[(iv)]
\begin{equation}
\mE_{\mSP}(R,U_{\cX}) = \sup_{\rho \in \bbR_+} \min_{q \in \cP(\cY)}\left\{ -\rho R -(1+\rho)\sum_{x \in \cX}U_{\cX}(x)\ln \sum_{y \in \cY}W(y|x)^{\frac{1}{1+\rho}}q(y)^{\frac{\rho}{1+\rho}}\right\},\label{eq:SP-dual}
\end{equation}
and $(\rho_R, q_R)$ is the unique saddle-point of \eqref{eq:SP-dual}.
\end{itemize}
\label{lem:SP-optimizers-new}
\end{proposition}
\vspace{-0.15cm}
\begin{IEEEproof}
The proof is given in Appendix~\ref{app-SP-opt}.\IEEEQEDoff
\end{IEEEproof}

Next, we state a concentration result, which is proven in \cite[Lemma~5]{altug13-phd} and reproduced here for completeness. Although there are various bounds of this sort, the classical versions in probability theory literature are stated in asymptotic form. 

To state the result, let $\{ Z_n\}_{n =1}^N$ be independent, real-valued random variables with law $\nu_n$, and assume
\begin{equation}
\sum_{n=1}^N \mV_{\nu_n}[Z_n] >0.	
\end{equation}
Define $\Lambda_n(\lambda) \eqdef \ln \mE_{\nu_n}\left[\textnormal{e}^{\lambda Z_n}\right]$ and assume the existence of a $c \in \bbR$ with a corresponding $\eta >0$ satisfying:
\begin{itemize}
\item[(i)] There exists a neighborhood of $\eta$ such that $\frac{1}{N}\sum_{n=1}^N \Lambda_n(\lambda) < \infty$, for all $\lambda$ in this neighborhood.
\item[(ii)] $\frac{1}{N}\sum_{n=1}^N \Lambda_n^\prime(\eta) = c$.
\end{itemize}
For any $b \in \bbR$, $\Lambda_N^\ast(b)$ denotes the Fenchel-Legendre transform of $\frac{1}{N} \sum_{n=1}^N \Lambda_n(\cdot)$ at $b$, i.e.,
\begin{equation}
\Lambda_N^\ast(b) \eqdef \sup_{\lambda \in \bbR}\left\{ \lambda b - \frac{1}{N} \sum_{n=1}^N \Lambda_n(\lambda) \right\}.
\end{equation}
Define
\begin{align}
\frac{\textnormal{d} \tilde{\nu}_{n}}{\textnormal{d} \nu_n} (z) & \eqdef \me^{\eta z - \Lambda_n(\eta)}, \\
T_n & \eqdef Z_n - \mE_{\tilde{\nu}_{n}}[Z_n], \\
m_{2,N} & \eqdef \sum_{n=1}^N \mV_{\tilde{\nu}_{n}}[T_n ], \\
m_{3,N} & \eqdef \sum_{n=1}^N \mE_{\tilde{\nu}_{n}}[|T_n|^3], \\
t_N & \eqdef \eta 2 \sqrt{2 \pi} \frac{ m_{3,N}}{m_{2,N}}.
\end{align}
\begin{lemma}
For any $N \in \bbZ^+$ and $a >1$,
\begin{equation}
\Pr \left[ \frac{1}{N} \sum_{n=1}^N Z_n \geq c \right]  \geq \me^{-a t_N}\left( 1 - \tfrac{1}{a}\right)(1+ a t_N)  \left\{ 1 - \tfrac{[1+(1+ a t_N)^2]}{(1+ a t_N) \eta\left(1 - 1/a\right)2 \sqrt{\me m_{2,N}}}\right\}\frac{1}{\eta\sqrt{2 \pi m_{2,N}}}\exp\left\{-N \Lambda_N^{\ast}(c)\right\}. \label{eq:EA-lower-bound}
\end{equation}
\label{lem:EA}
\end{lemma}
\begin{IEEEproof}
For completeness, we provide an outline of the proof in Appendix~\ref{app-EA}.\IEEEQEDoff
\end{IEEEproof}

We continue with a simple result for sums of independent random variables, which is used in the proofs of both Theorem~\ref{thrm:sym} and Theorem~\ref{thrm:asym}. Its derivation is inspired by the proof of \cite[Lemma~47]{polyanskiy10}; it is tighter than that result by at least a factor of $2$.
\begin{lemma}
Let $\{ Z_n\}_{n=1}^N$ be independent with
\begin{align}
m_{2,N} & \eqdef \sum_{n=1}^N \mV[Z_n] >0,\\
m_{3,N} & \eqdef \sum_{n=1}^N \mE\left[ | Z_n - \mE\left[ Z_n\right]|^3 \right] < \infty. 	
\end{align}
Then, for any $r \in \bbR$,
\begin{equation}
\mE\left[   \b1\left\{ \sum_{n=1}^N Z_n  \leq r \right\} \exp\left\{-\left[ r - \sum_{n=1}^N Z_n\right] \right\}\right] \leq \frac{1}{\sqrt{2 \pi m_{2,N}}} + \frac{2 m_{3,N}}{m_{2,N}^{3/2}}. \label{eq:lem2-1}
\end{equation}
Further, if the random variables are also identically distributed, then
\begin{equation}
\mE\left[   \b1\left\{ \sum_{n=1}^N Z_n  \leq r \right\} \exp\left\{-\left[ r - \sum_{n=1}^N Z_n\right] \right\}\right] \leq \frac{1}{\sqrt{2 \pi m_{2,N}}} + \frac{m_{3,N}}{m_{2,N}^{3/2}}. \label{eq:lem2-2}
\end{equation}
\label{lem:lem2}
\end{lemma}
\begin{IEEEproof}
The proof is given in Appendix~\ref{app-lem2}. \IEEEQEDoff
\end{IEEEproof}

\subsection{Proof of Theorem~\ref{thrm:nonsingular}}
\label{ssec:nonsingular}
The upper bound, \eqref{eq:nonsingular-ach}, follows from an application of \cite[Theorem~2(ii)]{altug13-a} with the pair $(U_{\cX},W)$, which is
nonsingular under \cite[Definition 1]{altug13-a} by 
Definition~\ref{def:singularity}.

To prove \eqref{eq:nonsingular-conv}, let $(f,\varphi)$ denote an arbitrary $(N,R)$ code with ideal feedback, and $\rho_R$ (resp. $q_R$) be as defined in \eqref{eq:rhoR} (resp. \eqref{eq:qR}). Evidently, $q_R(y)>0$ for all $y \in \cY$, since without loss of generality we can assume that $W$ has no all-zero columns. For any $R_\infty < r \leq R$, we define
\begin{equation}
\me_{\mSP}(r,R) \eqdef \inf_{V \in \cP(\cY|\cX) \colon \mD(V\|q_R|U_{\cX}) \leq r} \mD(V\|W|U_{\cX}).
\label{eq:eSP-nonsingular}
\end{equation}
For any $\bx^N \in \cX^N$, $m \in \cM$ and $r \in \bbR_+$, let
\begin{align}
\cS\left(\bx^N,r\right) & \eqdef \left\{\by^N \in \cY^N \colon \frac{1}{N}\sum_{n=1}^N\ln \frac{W(y_n|x_n)}{q_R(y_n)} \leq r - \me_{\mSP}(r,R) \right\}, \label{eq:SN-nonsingular}\\
 \cS(m,r) & \eqdef \left\{ \by^N \in \cY^N \colon \frac{1}{N}\sum_{n=1}^N \ln \frac{W(y_n|f_{n}(m,\by^{n-1}))}{q_R(y_n)} \leq r - \me_{\mSP}(r,R)\right\}. \label{eq:SN-nonsingular-fb}
\end{align}
We also use the notation $\cS\left(\bx^N,r\right)$ and $\cS(m,r)$ to refer to
the events
\begin{align*}
& \left\{ \bY^N \in \cS\left(\bx^N,r\right)\right\} \\
& \left\{ \bY^N \in \cS(m,r)\right\}.
\end{align*}
This convention will be used with
other similar quantities that are introduced later.

\begin{lemma}
\begin{itemize}
\item[(i)] For any $\lambda \in \bbR$, $M_{x}(\lambda) \eqdef  \sum_{y \in \supp(W(\cdot|x))}W(y|x)^{1-\lambda}q_R(y)^\lambda$ is finite and constant in $x \in \cX$.
\item[(ii)] For any $m \in \cM$ and $r \in \bbR_+$, $P_{\bY^N | M}\left\{ \cS(m,r) \, \big| \, m\right\} = W\left\{ \cS(\bx^N_\mo, r) \, \big| \, \bx_\mo^N\right\}$, where $\bx_\mo^N$ is an $N$-tuple consisting of all $x_\mo \in \cX$ and the choice of $x_\mo$ is immaterial in what follows.
\end{itemize}
\label{lem:iid-nonsingular}
\end{lemma}
\begin{IEEEproof}
\begin{itemize}
\item[(i)] $M_x(\lambda) \in \bbR$ directly follows from the fact that $W(\cdot|x) \ll q_R $ for any $x \in \cX$, which is a direct consequence of the fact that $\supp(q_R) = \cY$. Let $\{ \cY_l\}_{l=1}^L$ be a partition of the columns of $W$ mentioned in Definition~\ref{def:symmetric}, whose choice is immaterial in what follows. Since each column is a permutation of any other column for any sub-channel defined by this partition,
\begin{equation}
\left( \sum_{x \in \cX}U_{\cX}(x)W(y|x)^{\frac{1}{1+\rho_R}}\right)^{1+\rho_R}
\end{equation}
has the same value for any $y \in \cY_l$. This observation, coupled with the fact that every row is a permutation of every other row for any sub-channel defined by the aforementioned partition, suffices to conclude the proof of the second assertion.
\item[(ii)] For any $\lambda \in \bbR$, define
\begin{equation}
M_{m}(\lambda) \eqdef \mE_{P_{\bY^N|M}(\cdot|m)}\left[\exp\left\{\lambda \ln \tfrac{q_R(\bY^N)}{ P_{\bY^N|M}(\bY^N|m)}\right\}\right],	
\end{equation}
where $q_R(\by^N) \eqdef \prod_{n=1}^N q_R(y_n)$. We have
\begin{align}
M_{m}(\lambda) & =  \sum_{y_1 \in \cY} \ldots \sum_{y_N \in \cY}\prod_{n=1}^N W(y_n|f_{n}(m,\by^{n-1}))\exp\left\{\lambda \ln \frac{q_R(y_n)}{W(y_n|f_{n}(m,\by^{n-1}))}\right\} \\
& = [M_{x_{\mo}}(\lambda)]^N, \label{eq:lem-iid-nonsingular-pf1}
\end{align}
where \eqref{eq:lem-iid-nonsingular-pf1} follows from the first assertion of this lemma. Since
\begin{equation}
\mE_{W(\cdot|\bx^N_{\mo})}\left[\exp\left\{\lambda \ln  \frac{q_R(\bY^N)}{W(\bY^N|\bx^N_{\mo})}\right\}\right] =  [M_{x_{\mo}}(\lambda)]^N,	
\end{equation}
\eqref{eq:lem-iid-nonsingular-pf1} and the uniqueness theorem for the moment generating function (e.g., \cite[Ex.~26.7]{billingsley95}) imply the claim.
\end{itemize}
\end{IEEEproof}
For any $\lambda \in \bbR$, we define
\begin{equation}
\Lambda(\lambda) \eqdef \ln \mE_{W(\cdot|x_\mo)}\left[\exp\left\{\lambda \ln \frac{q_R(Y)}{W(Y|x_\mo)}\right\}\right]. \label{eq:Lambda-nonsingular}
\end{equation}
As a consequence of Lemma~\ref{lem:iid-nonsingular}(i), $\Lambda(\cdot)$ is
finite over the entire real line, which, in turn, ensures that $\Lambda(\cdot)$ is a smooth function on $\bbR$~\cite[Ex.~2.2.24]{dembo-zeitouni98}. For any $x \in \cX$, let
\begin{equation}
W_R(y|x) \eqdef \frac{q_R(y)}{q_R(\supp(W(\cdot|x)))}\b1\left\{y \in \supp(W(\cdot|x))\right\}.
\label{eq:WR-nonsingular}	
\end{equation}
Evidently, $W_R(\cdot|x) \equiv W(\cdot|x)$ for all $x \in \cX$. For any $x \in \cX$ and $\lambda \in [0,1)$, define
\begin{equation}
\tilde{W}_\lambda(y|x) \eqdef \frac{W(y|x)^{1-\lambda}q_R(y)^\lambda}{\sum_{b \in \cY} W(b|x)^{1-\lambda}q_R(b)^\lambda}. \label{eq:tiltedW-nonsingular}
\end{equation}
Via routine calculations, we deduce that
\begin{align}
\Lambda^\prime(\lambda) & = \mE_{\tilde{W}_\lambda(\cdot|x_\mo)}\left[ \ln \frac{q_R(Y)}{W(Y|x_\mo)}\right], \label{eq:moments-nonsingular-1} \\
\Lambda^{\prime \prime}(\lambda) & = \mV_{\tilde{W}_\lambda(\cdot|x_\mo)}\left[ \ln \frac{q_R(Y)}{W(Y|x_\mo)}\right]. \label{eq:moments-nonsingular-2}
\end{align}
Similarly, for any $\lambda \in [0,1)$, define
\begin{equation}
m_3(\lambda) \eqdef \mE_{\tilde{W}_\lambda(\cdot|x_\mo)}\left[ \left| \ln \frac{q_R(Y)}{W(Y|x_\mo)} - \Lambda^\prime(\lambda)\right|^3\right]. \label{eq:3rd-moment-nonsingular}
\end{equation}
From \eqref{eq:tiltedW-nonsingular}--\eqref{eq:3rd-moment-nonsingular}, one can verify that $\Lambda^\prime(\cdot), \Lambda^{\prime \prime}(\cdot)$ and $m_3(\cdot)$ are continuous over $[0,1)$. For any $b \in \bbR$, let $\Lambda^\ast(b)$ denote the Fenchel-Legendre transform of $\Lambda(\cdot)$ at $b$, i.e.,
\begin{equation}
\Lambda^\ast(b) = \sup_{\lambda \in \bbR}\left\{ \lambda b - \Lambda(\lambda)\right\}.
\label{eq:FL-nonsingular}
\end{equation}
The next result collects useful properties of the aforementioned quantities.
\begin{lemma}
\begin{itemize}
\item[(i)] $R > \mD(W_R\|q_R|U_\cX)$.

\item[(ii)] $\me_{\mSP}(R,R) = \mE_{\mSP}(R)$.

\item[(iii)] $\Lambda^{\prime \prime}(\lambda) >0$, for any $\lambda \in [0,1)$.

\item[(iv)] $s_{(\cdot)}: (\mD(W_R\|q_R|U_{\cX}), R] \to \bbR$ s.t. $s_r \eqdef -\left.\frac{\partial \me_{\mSP}(a,R)}{\partial a}\right|_{a = r}$ is a well-defined, continuous, positive and strictly decreasing function.

\item[(v)] Fix some $r \in (\mD(W_R\|q_R|U_{\cX}), R]$. We have 
\begin{equation}
\Lambda^\ast(\me_{\mSP}(r,R) - r ) = \me_{\mSP}(r,R). 	
\end{equation}
Moreover, $\eta_r \eqdef \frac{s_r}{1+s_r} \in (0,1)$ is the unique real number that satisfies
\begin{equation}
\Lambda^\prime(\eta_r) = \me_{\mSP}(r,R) - r. 	
\end{equation}
\item [(vi)]
$s_R = \rho_R$.
\end{itemize}
\label{lem:regularity-nonsingular}
\end{lemma}

\begin{IEEEproof}
The proof is given in Appendix~\ref{app-regularity-singular}.\IEEEQEDoff
\end{IEEEproof}
Define $\bar{R} \eqdef \frac{1}{2}(R + \mD(W_R\|q_R|U_{\cX}))$. Due to Lemma~\ref{lem:regularity-nonsingular}(i), $\bar{R} \in (\mD(W_R\|q_R|U_{\cX}), R)$. Moreover, as a direct consequence of Lemma~\ref{lem:regularity-nonsingular}(iv) and (v),
\begin{equation}
0 < \eta_{R} < \eta_r < \eta_{\bar{R}} < 1,
\end{equation}
for any $r \in (\bar{R}, R)$. Fix an arbitrary $a >1$ and define
\begin{align}
t_{\max} & \eqdef a 2 \sqrt{2 \pi} \eta_{\bar{R}} \max_{\lambda \in [0,\eta_{\bar{R}}]}\frac{m_3(\lambda)}{\Lambda^{\prime \prime}(\lambda)}, \\
m_{2,\min} & \eqdef \min_{\lambda\in [0,\eta_{\bar{R}}]} \Lambda^{\prime \prime}(\lambda), \\
m_{2,\max} & \eqdef \max_{\lambda\in [0,\eta_{\bar{R}}]} \Lambda^{\prime \prime}(\lambda).
\end{align}
Evidently all of the aforementioned quantities are well-defined, positive and finite. Finally, define
\begin{equation}
\frac{\me^{-t_{\max}}\left(1-\tfrac{1}{a} \right)}{\eta_{\bar{R}}2\sqrt{2 \pi m_{2,\max}}}=\mathrel{\mathop:}  k_\mo \in \bbR^+. \label{eq:N-large-nonsingular-01}
\end{equation}
Fix $k_1, k_2 \in \bbR^+$ that satisfy $k_2 - k_1 = \ln k_\mo$. Consider any $k_3 \in (0,1)$ that satisfies $\me^{-k_2}<k_3$. For any $N \in \bbZ^+$, define $R_N \eqdef R - \frac{1}{N}(k_1 + \ln\sqrt{N})$. Consider a sufficiently large $N \in \bbZ^+$, such that
\begin{align}
R_N & \geq \bar{R}, \label{eq:N-large-nonsingular-0}\\
\frac{1+(1+t_{\max})^2}{\eta_R \left(1 - 1/a\right) 2 \sqrt{\me N m_{2,\min}}}& \leq 1/2.
\label{eq:N-large-nonsingular-00}
\end{align}
For any $m \in \cM$, we have 
\begin{align}
P_{\bY^N|M}\left\{ \cS(m, R_N) | m \right\} & = W\left\{ \cS(\bx^N_\mo, R_N) | \bx^N_\mo \right\} \label{eq:bound-nonsingular0} \\
& \geq \frac{k_\mo}{\sqrt{N}}\exp\left\{-N \me_{\mSP}(R_N,R)\right\} \label{eq:bound-nonsingular02} \\
& > 0, \label{eq:bound-nonsingular1}
\end{align}
where \eqref{eq:bound-nonsingular0} follows from Lemma~\ref{lem:iid-nonsingular}(ii), \eqref{eq:bound-nonsingular02} follows from Lemma~\ref{lem:EA}, whose application is ensured by Lemma~\ref{lem:regularity-nonsingular}(iii) and (v), coupled with \eqref{eq:N-large-nonsingular-01}, \eqref{eq:N-large-nonsingular-0} and \eqref{eq:N-large-nonsingular-00}. By recalling \eqref{eq:Pe-nonsingular}, we continue as follows:
\begin{align}
\bar{\mP}_{\me}(f_N, \varphi_N) & \geq  \frac{1}{|\cM|}\sum_{m \in \cM} P_{\bY^N|M}\left\{ \cS(m,R_N) | m \right\} \sum_{\by^N \in \cA_m^c \cap \cS(m, R_N)}\frac{P_{\bY^N|M}(\by^N | m)}{P_{\bY^N|M}\left\{ \cS(m,R_N) | m \right\}}  \\
& \geq \frac{k_\mo}{\sqrt{N}}\exp\left\{-N \me_{\mSP}(R_N,R)\right\} \frac{1}{|\cM|}\sum_{m \in \cM}  \sum_{\by^N \in \cA_m^c \cap \cS(m, R_N)}\frac{P_{\bY^N|M}(\by^N | m)}{P_{\bY^N|M}\left\{ \cS(m,R_N) | m \right\}}, \label{eq:bound-nonsingular2}
\end{align}
where \eqref{eq:bound-nonsingular2} follows from \eqref{eq:bound-nonsingular02}. For any $m \in \cM$, we define
\begin{align}
P_{\bY^N|M,\cS(m,R_N)}(\by^N | m) & \eqdef \frac{P_{\bY^N|M}(\by^N|m)}{P_{\bY^N|M}\left\{ \cS(m, R_N) | m \right\}}\b1\left\{ \by^N \in \cS(m, R_N) \right\} , \label{eq:P_YX-nonsingular} \\
P_{\bY^N|\cS(m,R_N)}(\by^N) & \eqdef \frac{q_R(\by^N)}{q_R\left\{ \cS(m, R_N) \right\}}\b1\left\{ \by^N \in \cS(m, R_N)\right\}, \label{eq:P_Y-nonsingular}
\end{align}
and note that since $q_R \gg W(\cdot|x)$, \eqref{eq:bound-nonsingular1} ensures that both \eqref{eq:P_YX-nonsingular} and \eqref{eq:P_Y-nonsingular} are well-defined probability measures. By substituting \eqref{eq:P_YX-nonsingular} into \eqref{eq:bound-nonsingular2}, we deduce that
\begin{align}
\bar{\mP}_{\me}(f_N, \varphi_N)  & \geq \frac{k_\mo}{\sqrt{N}}\exp\left\{-N \me_{\mSP}(R_N,R)\right\} \frac{1}{|\cM|}\sum_{m \in \cM}  \sum_{\by^N \in \cA_m^c }P_{\bY^N|M,\cS(m,R_N)}(\by^N | m)  \\
& = \frac{k_\mo}{\sqrt{N}}\exp\left\{-N \me_{\mSP}(R_N,R)\right\} \left( 1 - \frac{1}{|\cM|}\sum_{m \in \cM}  \sum_{\by^N \in \cA_m }P_{\bY^N|M,\cS(m,R_N)}(\by^N | m)  \right). \label{eq:bound-nonsingular3}
\end{align}
We proceed with the following two lemmas:
\begin{lemma} For any $m \in \cM$,
\begin{equation}
\frac{1}{N}\ln \frac{P_{\bY^N|M,\cS(m,R_N)}(\by^N | m)}{P_{\bY^N|\cS(m,R_N)}(\by^N)} \leq R -\frac{k_2}{N},
\label{eq:emptyset-nonsingular}
\end{equation}
for all $\by^N \in \cY^N$ with $P_{\bY^N|M,\cS\left(m, R_N\right)}(\by^N|m)>0$.
\label{lem:emptyset-nonsingular}
\end{lemma}
\begin{IEEEproof}
Fix any $m \in \cM$ and $\by^N \in \cS\left(m, R_N\right)$ with $P_{\bY^N|M}(\by^N|m) >0$. We have
\begin{align}
\frac{1}{N}\ln \frac{P_{\bY^N|M,\cS(m,R_N)}(\by^N | m)}{P_{\bY^N|\cS(m,R_N)}(\by^N)} & = \frac{1}{N}\ln\frac{P_{\bY^N|M}(\by^N|m)}{q_R(\by^N)} + \frac{1}{N}\ln\frac{q_R\left\{ \cS(m,R_N)\right\}}{P_{\bY^N|M}\left\{ \cS(m,R_N) | m\right\}} \label{eq:lem-emptyset-nonsingular-pf1}\\
& \leq \frac{1}{N}\ln\frac{P_{\bY^N|M}(\by^N|m)}{q_R(\by^N)} + \me_{\mSP}(R_N,R) + \frac{\ln \sqrt{N}}{N} - \frac{\ln k_\mo}{N} \label{eq:lem-emptyset-nonsingular-pf2}\\
& \leq R - \frac{k_2}{N}, \label{eq:lem-emptyset-nonsingular-pf3}
\end{align}
where \eqref{eq:lem-emptyset-nonsingular-pf1} follows from the definitions of $P_{\bY^N|M,\cS(m,R_N)}$ and $P_{\bY^N|\cS(m,R_N)}$, i.e., \eqref{eq:P_YX-nonsingular} and \eqref{eq:P_Y-nonsingular}, \eqref{eq:lem-emptyset-nonsingular-pf2} follows from \eqref{eq:bound-nonsingular02} and \eqref{eq:lem-emptyset-nonsingular-pf3} follows from the definition of $\cS\left(m, R_N \right)$, i.e., \eqref{eq:SN-nonsingular}, along with the fact that $k_2 - k_1 =\ln k_\mo$.
\end{IEEEproof}
\begin{lemma}
For any $\{ \psi_n : \cM \times \cY^{n-1} \to \cX \}_{n=1}^N$ and $r \in (\mD(W_R \| q_R | U_{\cX}),R]$,
\begin{equation}
q_R\left\{ \frac{1}{N}\sum_{n=1}^N \ln \frac{W(Y_n | \psi_n(m,\bY^{n-1}))}{q_R(Y_n)} \leq r - \me_{\mSP}(r,R)\right\} \geq k_3, \label{eq:nonsingular-star}
\end{equation}
for all sufficiently large $N \in \bbZ^+$, independent of $m \in \cM$, where $k_3$ is defined right after \eqref{eq:N-large-nonsingular-01}.
\label{lem:nonsingular-star-lem}
\end{lemma}
\begin{IEEEproof}
Let $x_\mo \in \cX$ be as in Lemma~\ref{lem:iid-nonsingular}. First, note that
\begin{equation}
q_R\left\{ \frac{1}{N}\sum_{n=1}^N \ln \frac{W(Y_n|\psi_n(m,\bY^{n-1}))}{q_R(Y_n)} \leq r - \me_{\mSP}(r,R)\right\} = q_R\left\{ \frac{1}{N}\sum_{n=1}^N \ln \frac{W(Y_n|x_{\mo})}{q_R(Y_n)} \leq r - \me_{\mSP}(r,R) \right\}, \label{eq:nonsingular-star-pf1}
\end{equation}
which follows from the fact that, by the symmetry of the channel, for any $x \in \cX$, $\ln \frac{W(Y|x)}{q_R(Y)}$ and $\ln \frac{W(Y|x_\mo)}{q_R(Y)}$ have the same distribution when $Y$ has distribution $q_R$.

We conclude the proof of Theorem~\ref{thrm:nonsingular} as follows: first, assume that there exists a pair $(x,y) \in \cX \times \cY$ with $W(y|x)=0$. The symmetry of the channel ensures that there exists $y_\mo \in \cY$ such that $W(y_\mo|x_\mo) = 0$. Note that
\begin{align}
\left\{ \by^N \in \cY^N \colon \frac{1}{N}\sum_{n=1}^N \ln \frac{W(y_n|x_\mo)}{q_R(y_n)} > r - \me_{\mSP}(r,R) \right\} & \subseteq \{ \cY - \{ y_\mo\}\}^N, \label{eq:nonsingular-star-pf1-new-1} \\
q_R\{ \cY - \{ y_\mo\}\} & <1, \label{eq:nonsingular-star-pf1-new-2}
\end{align}
which are direct consequences of the fact that $\supp(q_R)=\cY$. From \eqref{eq:nonsingular-star-pf1-new-1} and \eqref{eq:nonsingular-star-pf1-new-2}, we conclude that
\begin{equation}
q_R\left\{ \frac{1}{N}\sum_{n=1}^N \ln \frac{W(Y_n|x_{\mo})}{q_R(Y_n)} \leq r - \me_{\mSP}(r,R) \right\} \geq k_3, \label{eq:nonsingular-star-pf2}
\end{equation}
for all sufficiently large $N \in \bbZ^+$.

Next, assume that for all $(x,y) \in \cX \times \cY$, $W(y|x)>0$. For any $\lambda \in \bbR$,
\begin{equation}
\Lambda_1(\lambda) \eqdef \ln \mE_{q_R}\left[\exp\left\{\lambda \ln \frac{W(Y|x_\mo)}{q_R(Y)}\right\}\right] = \Lambda(1-\lambda), \label{eq:nonsingular-star-pf3.0}
\end{equation}
as a direct consequence of the positivity of $W$. Equation~\eqref{eq:nonsingular-star-pf3.0}, along with Lemma~\ref{lem:regularity-nonsingular}(v), implies that there exists $\eta_r \in (0,1)$ with
\begin{equation}
\left[ \Lambda^\prime(\eta_r) = \me_{\mSP}(r,R)-r \right]\Longleftrightarrow \left[ \Lambda_1^{\prime}(1-\eta_r) = r - \me_{\mSP}(r,R) \right]. \label{eq:nonsingular-star-pf3.1}
\end{equation}
Further, Lemma~\ref{lem:regularity-nonsingular}(iii) ensures that
\begin{equation}
\left[\Lambda^{\prime \prime}(\cdot) >0 \right]\Longleftrightarrow \left[\Lambda_1^{\prime \prime}(1 - (\cdot)) >0 \right] \Longleftrightarrow \left[\Lambda_1^{\prime \prime}(\cdot) >0\right]. \label{eq:nonsingular-star-pf3.2}
\end{equation}
From \eqref{eq:nonsingular-star-pf3.1} and \eqref{eq:nonsingular-star-pf3.2}, we infer that
\begin{align}
\mu_{x_\mo} & \eqdef \mE_{q_R}\left[ \ln \frac{W(Y|x_\mo)}{q_R(Y)}\right] \\
& = \Lambda_1^\prime(0) \\
& < \Lambda_1^{\prime}(1-\eta_r) \\
& = r - \me_{\mSP}(r,R), \label{eq:nonsingular-star-pf3.3}\\
\sigma_{x_{\mo}}^2 & \eqdef \mV_{q_R}\left[ \ln \frac{W(Y|x_\mo)}{q_R(Y)} \right] \\
& = \Lambda_1^{\prime \prime}(0)  \in \bbR^+, \label{eq:nonsingular-star-pf3.4}
\end{align}
where the boundedness of $\Lambda_1^{\prime \prime}(0)$ is an immediate consequence of the positivity of $W$ and the fact that the input and output alphabets are finite. Hence, Chebyshev's inequality, coupled with \eqref{eq:nonsingular-star-pf3.3} and \eqref{eq:nonsingular-star-pf3.4}, implies that
\begin{equation}
q_R\left\{ \frac{1}{N}\sum_{n=1}^N \ln \frac{W(Y_n|x_{\mo})}{q_R(Y_n)} \leq r - \me_{\mSP}(r,R) \right\} \geq 1 - \frac{\sigma_{x_\mo}^2}{N [\Lambda_1^\prime(1-\eta_r) - \mu_{x_\mo}]^2} \geq k_3, \label{eq:nonsingular-star-pf3}
\end{equation}
for all sufficiently large $N \in \bbZ^+$. Equations~\eqref{eq:nonsingular-star-pf1}, \eqref{eq:nonsingular-star-pf2} and \eqref{eq:nonsingular-star-pf3} imply \eqref{eq:nonsingular-star}.
\end{IEEEproof}

By using Lemmas~\ref{lem:emptyset-nonsingular} and \ref{lem:nonsingular-star-lem}, along with the fact that the decoding regions are disjoint and $q_R$ is a probability measure, \eqref{eq:bound-nonsingular3} further implies that
\begin{equation}
\bar{\mP}_{\me}(f_N, \varphi_N)   \geq \left(1 - \frac{\me^{-k_2}}{k_3}\right) \frac{k_\mo}{\sqrt{N}}\exp\left\{-N \me_{\mSP}(R_N,R)\right\}. \label{eq:bound-nonsingular4}
\end{equation}

\begin{lemma}
Let $\varepsilon_N \eqdef \frac{ k_1 + \ln \sqrt{N}}{N}$.
\begin{equation}
\me_{\mSP}(R_N,R)  \leq \mE_{\mSP}(R) + \varepsilon_N |\mE_{\mSP}^\prime(R)| + \varepsilon_N^2 \frac{(1+s_{\bar{R}})^2}{2 m_{2,\min}}(1+|\mE_{\mSP}^\prime(R)|). \label{eq:exponent-nonsingular}
\end{equation}
\label{lem:exponent-nonsingular}
\end{lemma}
\vspace{-0.5cm}
\begin{IEEEproof}
The proof is given in Appendix~\ref{app-exponent-nonsingular}.\IEEEQEDoff
\end{IEEEproof}
Let $N \in \bbZ^+$ be sufficiently large such that
\begin{equation}
\exp\left\{-N \varepsilon_N^2 \frac{(1+s_{\bar{R}})^2}{2 m_{2,\min}}(1+|\mE_{\mSP}^\prime(R)|)\right\}\geq \frac{1}{2}.	
\end{equation}
Then, Lemma~\ref{lem:exponent-nonsingular} and \eqref{eq:bound-nonsingular4} imply that
\begin{equation}
\bar{\mP}_{\me}(f_N, \varphi_N) \geq \frac{k_\mo}{2}\left(1-\frac{\me^{-k_2}}{k_3}\right)\exp\left\{-k_1|\mE_{\mSP}^\prime(R)|\right\}\frac{\exp\left\{-N\mE_{\mSP}(R)\right\}}{N^{\frac{1}{2}(1+|\mE_{\mSP}^\prime(R)|)}}. \label{eq:bound-nonsingular5}
\end{equation}
Since the code is arbitrary, \eqref{eq:bound-nonsingular5} implies \eqref{eq:nonsingular-conv}.\hfill \IEEEQED

\subsection{Proof of Theorem~\ref{thrm:singular}}
\label{ssec:singular}
The achievability proof is similar to its counterpart in Theorem~\ref{thrm:nonsingular}. In particular, we begin by invoking \cite[Corollary~1(i)]{altug13-a} with the pair $(U_{\cX},W)$. However, in that result the singularity of the pairs in $\cP(\cX) \times \cP(\cY|\cX)$, which differs from the singularity of symmetric channels in Definition~\ref{def:singularity}, is the crucial assumption. As we note next, however, the fact that $W$ is a singular symmetric channel implies that the pair $(U_{\cX},W)$ is singular. Specifically, note that since $W \in \cP(\cY|\cX)$ is a singular symmetric channel, we have
\begin{equation}
\forall \, (x, y, z) \in \cX \times \cY \times \cX, \textnormal{ s.t. } U_{\cX}(x)	U_{\cX}(z)W(y|x)W(y|x) >0, W(y|x)=W(y|z),
\end{equation}
which, in light of \cite[Definition~1]{altug13-a}, ensures that the pair $(U_\cX,W)$ is singular. Owing to the symmetry of the channel, $\mE_{\mr}(\cdot,U_\cX) = \mE_{\mr}(\cdot)$ on $(R_\mcr,C(W))$ (e.g., \cite[p.~145]{gallager68}). Since $(U_\cX, W)$ pair is singular, \eqref{eq:singular-ach} is a direct consequence of \cite[Corollary~1(i)]{altug13-a}.

In order to prove the converse, let $(f_N, \varphi_N)$ denote an arbitrary $(N,R)$ code with ideal feedback, and recall that $q(y) \eqdef \sum_{x \in \cX}U_{\cX}(x)W(y|x)$. Due to the singularity of $W$, given any $y \in \cY$, $W(y|\cdot)$ is either zero or a positive constant that only depends on $y$, say $\xi_y$. Hence,
\begin{equation}
q(y) = \xi_y \alpha_y \, \textnormal{  with  } \, \alpha_y \eqdef \sum_{x:W(y|x)>0}U_{\cX}(x).
\label{eq:singular-conv1}
\end{equation}
Since, without loss of generality, we can assume that $W$ has no all-zero columns, $q(y) >0$ for all $y \in \cY$ and hence $q\gg W(\cdot|x)$ for any $x \in \cX$. For any $r \in \bbR_+$, define
\begin{align}
\cS(r) & \eqdef \left\{ \by^N \in \cY^N : \frac{1}{N} \sum_{n=1}^N \ln \frac{1}{\alpha_{y_n}} \leq r \right\} \label{eq:singular-conv2} \\
     & = \left\{ \by^N \in \cY^N : \frac{1}{N} \sum_{n=1}^N \ln \frac{W(y_i|x_i)}{q(y_n)} \leq r \quad \text{for some $\bx^N$ such that $W(\by^N|\bx^N) > 0$}\right\}.
\end{align}
Let $\bar{R} \eqdef \frac{R+R_\infty}{2}$. Fix some $k \in \bbR^+$ and define $R_N \eqdef R - \frac{k}{N}$. Consider a sufficiently large $N$, such that $R_N \geq \bar{R}$.
\begin{lemma}
Let $\bx^N_\mo$ denote the sequence consisting of $x_\mo \in \cX$ repeated $N$ times for some $x_\mo$, whose choice is immaterial in what follows. Consider any $\{\psi_n\}_{n=1}^N$ with $\psi_1 \in \cX$ and $\psi_n : \cY^{n-1} \to \cX$ for all $n \in \{2,\ldots, N\}$.
\begin{itemize}
\item[(i)] For any $r \in \bbR^+$,
\begin{equation}
\sum_{\by^N \in \cS(r)}W(y_1|\psi_1)\prod_{n=2}^N W(y_n | \psi_n(\by^{n-1})) = W\left\{ \cS(r) | \bx^N_{\mo}\right\}.	
\end{equation}
\item[(ii)]For some $\tilde{K} \in \bbR^+$ that depends on $R, \bar{R}$ and $W$,
\begin{equation}
W\left\{ \cS(R_N) | \bx^N_\mo \right\} \geq \frac{\tilde{K}}{\sqrt{N}}\exp\left\{-N \mE_{\mSP}(R)\right\} >0,	
\end{equation}
for all sufficiently large $N$.
\end{itemize}
\label{lem:singular-lem1}
\end{lemma}
\begin{IEEEproof}
The proof is given in Appendix~\ref{app-singular-lem1}.\IEEEQEDoff
\end{IEEEproof}
Similar to \eqref{eq:bound-nonsingular2}, from \eqref{eq:Pe-nonsingular}, along with Lemma~\ref{lem:singular-lem1}, we infer that
\begin{equation}
\bar{\mP}_{\me}(f_N, \varphi_N) \geq \frac{\tilde{K}}{\sqrt{N}}\exp\left\{-N \mE_{\mSP}(R)\right\} \frac{1}{|\cM|}\sum_{m \in \cM}\sum_{\by^N \in \cA_m^c \cap \cS(R_N)}\frac{P_{\bY^N|M}(\by^N|m)}{P_{\bY^N|M}\left\{ \cS(R_N)|m\right\}}. \label{eq:singular-conv3}
\end{equation}
For all $m \in \cM$, define
\begin{align}
P_{\bY^N|M,\cS(R_N)}(\by^N | m) & \eqdef \frac{P_{\bY^N|M}(\by^N|m)}{P_{\bY^N|M}\left\{ \cS(R_N) | m \right\}}\b1\left\{ \by^N \in \cS(R_N) \right\}, \label{eq:P_YX-singular} \\
P_{\bY^N|\cS(R_N)}(\by^N) & \eqdef \frac{q(\by^N)}{q\left\{ \cS(R_N) \right\}}\b1\left\{ \by^N \in \cS(R_N) \right\}. \label{eq:P_Y-singular}
\end{align}
Due to Lemma~\ref{lem:singular-lem1} and the fact that $q \gg W(\cdot|x)$, \eqref{eq:P_YX-singular} and \eqref{eq:P_Y-singular} are well-defined probability measures. By substituting \eqref{eq:P_YX-singular} in \eqref{eq:singular-conv3}, one can check that
\begin{equation}
\bar{\mP}_{\me}(f_N, \varphi_N) \geq \frac{\tilde{K}}{\sqrt{N}}\exp\left\{-N \mE_{\mSP}(R)\right\} \left( 1 - \frac{1}{|\cM|}\sum_{m \in \cM}\sum_{\by^N \in \cA_m} P_{\bY^N|M,\cS(R_N)}(\by^N | m) \right). \label{eq:singular-conv4}
\end{equation}

\begin{lemma}
For any $m \in \cM$,
\begin{equation}
\frac{1}{N}\ln \frac{P_{\bY^N|M,\cS(R_N)}(\by^N | m)}{P_{\bY^N|\cS(R_N)}(\by^N)} \leq R - \frac{k}{N},
\end{equation}
for all $\by^N \in \cY^N$ with $P_{\bY^N|M,\cS(R_N)}(\by^N | m) >0$.
\label{lem:singular-lem2}
\end{lemma}
\begin{IEEEproof}
Fix any $m \in \cM$ and $\by^N \in \cS(R_N)$ with $P_{\bY^N|M}(\by^N|m) >0$. First, we claim that
\begin{equation}
q(\cS(R_N)) = P_{\bY^N|M}\left\{ \cS(R_N) | m\right\}. \label{eq:singular-lem2-pf1}
\end{equation}
To see this, note that
\begin{align}
q(\cS(R_N))  & = \sum_{\bx^N \in \cX^N}U_{\cX^N}(\bx^N) \sum_{\by^N \in \cY^N}W(\by^N|\bx^N)\b1\left\{ \by^N \in \cS(R_N) \right\}  \\
& = \sum_{\bx^N \in \cX^N}U_{\cX^N}(\bx^N) W\left\{ \cS(R_N) | \bx^N \right\}  \\
& = \sum_{\bx^N \in \cX^N}U_{\cX^N}(\bx^N) W\left\{ \cS(R_N) | \bx_\mo^N \right\} \label{eq:singular-lem2-pf2} \\
& = P_{\bY^N|M}\left\{ \cS(R_N) | m \right\}, \label{eq:singular-lem2-pf3}
\end{align}
where \eqref{eq:singular-lem2-pf2} and \eqref{eq:singular-lem2-pf3} follow from Lemma~\ref{lem:singular-lem1}(i). Hence,
\begin{align}
\frac{1}{N}\ln \frac{P_{\bY^N|M,\cS(R_N)}(\by^N | m)}{P_{\bY^N|\cS(R_N)}(\by^N)} & = \frac{1}{N}\ln \frac{P_{\bY^N|M}(\by^N|m)}{q(\by^N)} \label{eq:singular-lem2-pf4}\\
& = \frac{1}{N}\sum_{n=1}^N \ln \frac{1}{\alpha_{y_n}} \label{eq:singular-lem2-pf5}\\
&  \leq R - \frac{k}{N}, \label{eq:singular-lem2-pf6}
\end{align}
 where \eqref{eq:singular-lem2-pf4} follows from \eqref{eq:singular-lem2-pf1}, \eqref{eq:singular-lem2-pf5} follows from the fact that whenever $W(y|x)>0$, $\frac{W(y|x)}{q(y)}= \frac{1}{\alpha_y}$, which is a direct consequence of the singularity of the channel, and \eqref{eq:singular-lem2-pf6} follows from the definition of $\cS(R_N)$, i.e., \eqref{eq:singular-conv2}.
\end{IEEEproof}

By using Lemma~\ref{lem:singular-lem2}, along with the fact that the decoding regions are disjoint and $P_{\bY^N|\cS(R_N)}$ is a probability measure, \eqref{eq:singular-conv4} implies that
\begin{equation}
\bar{\mP}_{\me}(f_N, \varphi_N) \geq \tilde{K}\left( 1 - \me^{-k}\right)\frac{1}{\sqrt{N}}\exp\left\{-N \mE_{\mSP}(R)\right\}.
\label{eq:singular-conv5}
\end{equation}
Since the code is arbitrary, \eqref{eq:singular-conv5} implies \eqref{eq:singular-conv}.\hfill \IEEEQED

\subsection{Proof of Theorem~\ref{thrm:sym}}
\label{ssec:sym}
Let $W \in \cP(\cY | \cX)$ be a symmetric and singular channel
with $V_\epsilon(W) > 0$. Without loss of generality, assume $W$ has no all-zero columns. Consider any $\epsilon \in (0,1)$. Similar to Section~\ref{ssec:singular}, define
\begin{equation}
\forall \, x \in \cX, \, M_x(\lambda) \eqdef \mE_{W(\cdot|x)}\left[ e^{\lambda \ln \frac{W(Y|x)}{q(Y)}}\right], \, m_3(x) \eqdef \mE_{W(\cdot|x)}\left[ \left| \ln \frac{W(Y|x)}{q(Y)} - C(W) \right|^3\right],
\end{equation}
for any $\lambda \in \bbR$ (recall that $q(\cdot)$ is the output
distribution induced by the uniform input distribution). In the proof to follow, we essentially use the same idea given in Section~\ref{ssec:singular}, and in particular the set $\cS(R)$, which is defined in \eqref{eq:singular-conv2}.

\begin{lemma} Let $W \in \cP(\cY|\cX)$ be a symmetric and singular channel.
Write $\alpha_y$ for $\alpha_y(U_{\cX})$.
Fix an arbitrary $x_\mo \in \cX$.
\begin{itemize}
\item[(i)] For any $x \in \cX$, $M_x(\lambda) = M_{x_{\mo}}(\lambda)$ for all $\lambda \in \bbR$.

\item[(ii)] For all $x \in \cX$,
\begin{align}
\mE_{W(\cdot|x)}\left[ \ln \frac{W(Y|x)}{q(Y)}\right] & = \mE_{W(\cdot|x_\mo)}\left[ \ln \frac{W(Y|x_\mo)}{q(Y)}\right]  \label{eq:lem-sym-prelim-1} \\
& = C(W), \label{eq:lem-sym-prelim-1.0}\\
\mV_{W(\cdot|x)}\left[ \ln \frac{W(Y|x)}{q(Y)}\right] & = \mV_{W(\cdot|x_\mo)}\left[ \ln \frac{W(Y|x_\mo)}{q(Y)}\right] \label{eq:lem-sym-prelim-1.1}\\
& =: V(W) \\
& = V_\epsilon(W), \label{eq:lem-sym-prelim-2}\\
m_3(x) & = m_3(x_\mo). \label{eq:lem-sym-prelim-3}
\end{align}

\item[(iii)] For any $m \in \cM$,
\begin{equation}
P_{\bY^N|M}\left\{ \cS(R) | m\right\} = W\left\{ \cS(R) | \bx^N_{\mo}\right\}.\label{em-sym-prelim-4}
\end{equation}

\item[(iv)] \begin{align}
 \mE_{q}[-\ln \alpha_Y] & = C(W),\\
 \mV_q[-\ln \alpha_Y] & = V(W), \\	
\mE_q[|-\ln \alpha_Y - C(W)|^3] & = m_3(x_\mo).
 \end{align}
\end{itemize}
\label{lem:sym-prelim}
\end{lemma}
\begin{IEEEproof}
Since $U_\cX$ is a capacity achieving input distribution of $W$ (e.g., \cite[Theorem~4.5.2]{gallager68}) and the unique capacity achieving output distribution has full support (e.g., \cite[Corollary 1 and 2 to Theorem~4.5.1]{gallager68}), we conclude that $\alpha_y >0$, for all $y \in \cY$.

\begin{itemize}
\item[(i)] The assertion has already been proven in the beginning of the proof of Lemma~\ref{lem:singular-lem1}, given in Appendix~\ref{app-singular-lem1}.

\item[(ii)] The first assertion of this lemma, along with the uniqueness theorem for the moment generating function (e.g., \cite[Ex.~26.7]{billingsley95}), directly implies \eqref{eq:lem-sym-prelim-1}, \eqref{eq:lem-sym-prelim-1.0}, \eqref{eq:lem-sym-prelim-1.1}, and \eqref{eq:lem-sym-prelim-3}. \eqref{eq:lem-sym-prelim-2} is evident in light of \eqref{eq:lem-sym-prelim-1.1} and the fact that $q$ is the unique capacity achieving output distribution of $W$.

\item[(iii)] The assertion is a direct consequence of Lemma~\ref{lem:singular-lem1}(i) by particularizing it to $\{\psi_n(\cdot)\}_{n=1}^N \leftarrow \{ f_n(m,\cdot)\}_{n=1}^N$ and $r \leftarrow R$.

\item[(iv)] The claim directly follows from the second assertion of this lemma on account of the definition of $q$ and the fact that $q(y) = \xi_y \alpha_y$. \hfill \IEEEQED
\end{itemize}
\IEEEQEDoff
\end{IEEEproof}

Returning to the proof of Theorem~\ref{thrm:sym}, we first define
\begin{align}
k(W) & \eqdef \frac{m_3(x_\mo)}{V(W)^{3/2}}, \\
K(\epsilon, W) & \eqdef \frac{k(W)\sqrt{V(W)}}{\phi(\Phi^{-1}(\epsilon))} + \frac{2}{\phi(\Phi^{-1}(\epsilon))}\left( \frac{1}{\sqrt{2\pi}} + \frac{m_3(x_\mo)}{V(W)}\right). \label{eq:sym-pf-final0}
\end{align}
Evidently, $K(\epsilon, W) \in \bbR^+$. Choose some $N_\mo(\epsilon, W) \in \bbZ^+$ such that for all $N \geq N_\mo(\epsilon,W)$,
\begin{equation}
1 - \frac{K(\epsilon,W)}{2 \phi(\Phi^{-1}(\epsilon))\sqrt{N \cdot V(W)}} > 1/2.
\label{eq:sym-pf-final1}
\end{equation}
Consider any $N \geq N_\mo(\epsilon,W)$ and define
\begin{equation}
R \eqdef C(W) + \sqrt{\frac{V(W)}{N}}\Phi^{-1}(\epsilon) + \frac{K(\epsilon,W)}{N}.
\label{eq:sym-pf-final2}
\end{equation}
Let $(f, \varphi)$ be an arbitrary $(N,R)$ code with feedback. We claim that
\begin{equation}
\bar{\mP}_{\me}(f, \varphi) \geq W\{\cS(R)|\bx_{\mo}^N\} - \sum_{\by^N \in \cS(R)}q(\by^N)\exp\left\{-N\left[R-\frac{1}{N}\sum_{k=1}^N\ln \frac{1}{\alpha_{y_k}}\right]\right\}, \label{eq:prop1-pf1}
\end{equation}
where $\bar{\mP}_{\me}(f, \varphi)$ denotes the average error probability of the code $(f, \varphi)$. To see \eqref{eq:prop1-pf1}, assume $W\{\cS(R)|\bx^N_{\mo}\} >0$, because otherwise \eqref{eq:prop1-pf1} is trivially true. Also, recall that $\cA_m \in \cY^n$ denotes the decoding region corresponding to the message $m \in \cM$. Define the following probability distributions
\begin{align}
P_{\bY^N|M,\cS(R)}(\by^N|m,\cS(R)) & \eqdef
\frac{P_{\bY^N|M}(\by^N|m)}{P_{\bY^N|M}\{\cS(R)|m\}}\b1\left\{ \by^N \in \cS(R)\right\}
\label{eq:PYMS}\\
P_{D|\bY^N}(m|\by^N) & \eqdef \b1\left\{ \by^N \in \cA_m \right\}, \label{eq:PDY}
\end{align}
and note that
\begin{align}
\bar{\mP}_{\me}(f, \varphi) & = \frac{1}{|\cM|}\sum_{m \in \cM}\sum_{\by^N \in \cA_m^c}P_{\bY^N|M}(\by^N|m) \\
& \geq \frac{1}{|\cM|}\sum_{m \in \cM}\sum_{\by^N \in \cA_m^c \cap \cS(R)}P_{\bY^N|M}(\by^N|m) \\
& = \frac{1}{|\cM|}\sum_{m \in \cM}P_{\bY^N|M}\{\cS(R)|m\}\sum_{\by^N \in \cA_m^c}P_{\bY^N|M,\cS(R)}(\by^N|m,\cS(R)) \\
& = W\{\cS(R)|\bx^N_{\mo}\}\left\{ 1 - \frac{1}{|\cM|}\sum_{m \in \cM}\sum_{\by^N \in \cA_m}P_{\bY^N|M,\cS(R)}(\by^N|m,\cS(R))\right\} \label{eq:prop1-pf2}\\
& = W\{\cS(R)|\bx^N_{\mo}\} \left\{ 1 - \frac{\me^{-NR}}{W\{\cS(R)|\bx^N_{\mo}\}} \sum_{m \in \cM}\sum_{\by^N \in \cY^N}P_{D|\bY^N}(m|\by^N)P_{\bY^N|M}(\by^N|m)\b1\left\{ \by^N \in \cS(R) \right\}\right\} \label{eq:prop1-pf3} \\
& \geq W\{\cS(R)|\bx^N_{\mo}\} \left\{ 1 - \frac{\me^{-NR}}{W\{\cS(R)|\bx^N_{\mo}\}} \sum_{m \in \cM}\sum_{\by^N \in \cS(R)}P_{D|\bY^N}(m|\by^N) q(\by^N)\exp\left\{\sum_{k=1}^N \ln \frac{1}{\alpha_{y_k}}\right\} \right\} \label{eq:prop1-pf4}\\
& \geq W\{\cS(R)|\bx^N_{\mo}\} - \sum_{\by^N \in \cS(R)}q(\by^N)\exp\left\{-N\left[ R - \frac{1}{N}\sum_{k=1}^N \ln \frac{1}{\alpha_{y_k}}\right]\right\}, \label{eq:prop1-pf5}
\end{align}
where in \eqref{eq:prop1-pf2} and \eqref{eq:prop1-pf3} we use Lemma~\ref{lem:sym-prelim}(iii), and \eqref{eq:prop1-pf4} follows from the fact that $q$ dominates $W(\cdot|x)$ for any $x \in \cX$, along with the singularity of the channel.
This establishes \eqref{eq:prop1-pf1}.

Since $V(W)>0$, Lemma~\ref{lem:sym-prelim}(iv) enables us to apply Lemma~\ref{lem:lem2} to deduce that
\begin{equation}
\sum_{\by^N \in \cS(R)}q(\by^N)\exp\left\{-N\left[R - \frac{1}{N}\sum_{i=1}^N \ln \frac{1}{\alpha_{y_i}}\right]\right\} \leq \frac{1}{\sqrt{2 \pi N \cdot V(W)}} + \frac{k(W)}{\sqrt{N}}. \label{eq:sym-pf-final4}
\end{equation}
Next, we claim that
\begin{equation}
W(\cS(R) | \bx^N_\mo) \geq \epsilon + \frac{K(\epsilon,W)\phi(\Phi^{-1}(\epsilon))}{\sqrt{N\cdot V(W)}}\left\{ 1 - \frac{K(\epsilon,W)}{\phi(\Phi^{-1}(\epsilon))2 \sqrt{N \cdot V(W)}}\right\} - \frac{k(W)}{2 \sqrt{N}}.
\label{eq:sym-pf-final5}
\end{equation}
To see \eqref{eq:sym-pf-final5}, we note that
\begin{align}
W(\cS(R) | \bx^N_\mo) & = W\left\{  \frac{1}{N} \sum_{i=1}^N \ln \frac{W(Y_i|x_\mo)}{q(Y_i)} \leq R \, \bigg| \, \bx^N_\mo \right\} \label{eq:sym-pf-final5.1} \\
& = W\left\{\frac{1}{\sqrt{N \cdot V(W)}}\sum_{i=1}^N \left[ \ln \frac{W(Y_i|x_\mo)}{q(Y_i)} - C(W) \right] \leq \Phi^{-1}(\epsilon) + \frac{K(\epsilon,W)}{\sqrt{N \cdot V(W)}} \, \bigg| \, \bx^N_\mo \right\} \label{eq:sym-pf-final5.2} \\
& \geq \Phi\left( \Phi^{-1}(\epsilon) + \frac{K(\epsilon, W)}{\sqrt{N \cdot V(W)}}\right) - \frac{k(W)}{2\sqrt{N}}, \label{eq:sym-pf-final5.3}
\end{align}
where \eqref{eq:sym-pf-final5.1} follows since $q(y) = \xi_y \alpha_y$, along with the singularity of the channel, \eqref{eq:sym-pf-final5.2} follows from the definition of $R$, i.e., \eqref{eq:sym-pf-final2}, and \eqref{eq:sym-pf-final5.3} follows from the Berry-Esseen Theorem\footnote{For convenience, we take the universal constant as $1$, although it is not the best possible for independent random variables. See \cite{korolev2010} for a survey on the constants of this theorem.}, whose applicability is ensured by Lemma~\ref{lem:sym-prelim}(ii) and the fact that $V(W)>0$. Via a second-order power series expansion, one can check that \eqref{eq:sym-pf-final5.3} implies \eqref{eq:sym-pf-final5}.

By substituting \eqref{eq:sym-pf-final4} and \eqref{eq:sym-pf-final5} into \eqref{eq:prop1-pf1}, along with \eqref{eq:sym-pf-final1} and noticing the fact that the code is arbitrary, we deduce that eventually,
\begin{equation}
\bar{\mP}_{\me}(N, R) > \epsilon,
\end{equation}
which implies that eventually,
\begin{equation}
\ln M_{\textnormal{fb}}^\ast(N, \epsilon) \leq N\cdot C(W) + \sqrt{N \cdot V(W)}\Phi^{-1}(\epsilon) + K(\epsilon, W),
\end{equation}
which, in turn, implies the desired result. \hfill \IEEEQED

\subsection{Proof of Theorem~\ref{thrm:asym}}
\label{ssec:asym}
For any $Q \in \cP(\cX)$, define
\begin{equation}
\alpha_y(Q) \eqdef \sum_{x:W(y|x)>0}Q(x),
\label{eq:alpha}
\end{equation}
and consider any singular $W \in \cP(\cY|\cX)$. As mentioned before, the singularity ensures that for any $y \in \cY$, $W(y|x)$ is either $0$ or a column-specific positive constant $\xi_y$. For any $y \in \cY$,
\begin{equation}
q_Q(y) = \xi_y \alpha_y(Q). 	
\end{equation}
The following set, which is a generalization of \eqref{eq:singular-conv2}, is instrumental in our analysis:
\begin{align}
\cS_R(Q) & \eqdef \left\{ \by^N : \frac{1}{N}\sum_{i=1}^N \ln \frac{1}{\alpha_{y_i}(Q)} \leq R \right\},
\label{eq:SR}
\end{align}
for any $R \in \bbR_+$.

\begin{lemma}
Consider a singular $W \in \cP(\cY|\cX)$. Consider any $(N,R)$ code, say $(f, \varphi)$, with codewords $\{ \bx^n(m)\}_{m=1}^{|\cM|}$. Let $\bar{\mP}_{\me}(f, \varphi)$ denote the average error probability of this code. Fix some $Q \in \cP(\cX)$ and $\bz^N \in \cX^N$ and assume that for all $m \in \cM$, $W(\cS_R(Q)|\bx^n(m)) = W(\cS_R(Q) | \bz^N)$ and $q_Q$ dominates $W(\cdot|x)$ for all $x \in \supp(P_{\bx^N(m)})$, where $P_{\bx^N(m)}$ denotes the empirical distribution of $\bx^N(m)$. Then,
\begin{equation}
\bar{\mP}_{\me}(f, \varphi) \geq W(\cS_R(Q) | \bz^N) - \sum_{\by^N \in \cS_R(Q)}q_Q(\by^N) \exp\left\{-N\left[ R - \frac{1}{N}\sum_{i=1}^N \ln \frac{1}{\alpha_{y_i}(Q)}\right]\right\}.
\label{eq:lem-multipurp}
\end{equation}
\label{lem:multipurp}
\end{lemma}
\vspace{-0.25cm}
\begin{IEEEproof}
Assume $W(\cS_R(Q)|\bz^N) >0$, otherwise \eqref{eq:lem-multipurp} is trivial. For any $\bx^N \in \cX^N$ with $W(\cS_R(Q) | \bx^N) >0$, define
\begin{equation}
P_{\bY^N|\bX^N, \cS_R(Q)}(\by^N | \bx^N , \cS_R(Q)) \eqdef \frac{W(\by^N | \bx^N )}{W( \cS_R(Q) | \bx^N)}\b1\{\by^N \in \cS_R(Q)\}.
\label{eq:lem-multipurp-cond-dist}
\end{equation}
Evidently, $P_{\bY^N|\bX^N, \cS_R(Q)}(\cdot | \bx^N , \cS_R(Q)) $ is a well-defined probability measure. As before, $\{ \mathcal{A}_m\}_{m=1}^{|\cM|}$ denote the decoding regions of the code and
\begin{align}
\bar{\mP}_{\me}(f, \varphi) & = \frac{1}{|\cM|}\sum_{m \in \cM} \sum_{\by^N \in \cA_m^c}W(\by^N | \bx^N(m))  \\
& \geq \frac{1}{|\cM|}\sum_{m \in \cM} \sum_{\by^N \in \cA_m^c} W(\cS_R(Q) | \bx^N(m)) P_{\bY^N|\bX^N, \cS_R(Q)}(\by^N | \bx^N(m), \cS_R(Q)) \label{eq:lem-multipurp-pf1} \\
& \geq W(\cS_R(Q) | \bz^N) \left[1 - \frac{1}{|\cM|}\sum_{m \in \cM} \sum_{\by^N \in \cA_m}  P_{\bY^N|\bX^N, \cS_R(Q)}(\by^N | \bx^N(m), \cS_R(Q)) \right]\label{eq:lem-multipurp-pf2},
\end{align}
where \eqref{eq:lem-multipurp-pf1} follows from \eqref{eq:lem-multipurp-cond-dist} and \eqref{eq:lem-multipurp-pf2} follows from the assumption that $W(\cS_R(Q)|\bx^N(m)) = W(\cS_R(Q) | \bz^N)$, for all $m \in \cM$. As before, define $P_{D|Y}(m| \by^N) \eqdef \b1\{ \by^N \in \cA_m\}$, for all $m \in \cM$. Since the decoding regions are mutually exclusive and collectively exhaustive on $\cM$, $P_{D|Y}(\cdot| \by^N) $ is a well-defined probability measure. Hence, \eqref{eq:lem-multipurp-pf2} implies that
\begin{align}
\bar{\mP}_{\me}(f, \varphi) & \geq W(\cS_R(Q) | \bz^N) \left[ 1 - \frac{\me^{-NR}}{W(\cS_R(Q)| \bz^N)} \sum_{m \in \cM} \sum_{\by^N} P_{D|Y}(m|\by^N) W(\by^N | \bx^N(m)) \b1\{ \by^N \in \cS_R(Q)\}\right]  \\
& \geq  W(\cS_R(Q) | \bz^N) \left[ 1 - \frac{\me^{-NR}}{W(\cS_R(Q)| \bz^N)} \sum_{m \in \cM} \sum_{\by^N} P_{D|Y}(m|\by^N)  \b1\{ \by^N \in \cS_R(Q)\} q_Q(\by^N)\me^{\sum_{i=1}^N \ln \frac{1}{\alpha_{y_i}(Q)}}  \right] \label{eq:lem-multipurp-pf3}\\
& \geq W(\cS_R(Q) | \bz^N) \left[ 1 - \frac{\me^{-NR}}{W(\cS_R(Q)| \bz^N)}  \sum_{\by^N} \b1\{ \by^N \in \cS_R(Q)\} q_Q(\by^N)\me^{\sum_{i=1}^N \ln \frac{1}{\alpha_{y_i}(Q)}}  \right],
\end{align}
where \eqref{eq:lem-multipurp-pf3} follows from the fact that $q_Q(y) = \xi_y \alpha_y(Q)$ and the assumption that for all $m \in \cM$, $q_Q$ dominates $W(\cdot|x)$ for all $x \in \supp(P_{\bx^N(m)})$.
\end{IEEEproof}
We analyze three different possibilities for the composition of the code $P$: large $\mI(P;W)$ with large $V(P,W)$, large $\mI(P;W)$ with small $V(P,W)$, and small $\mI(P;W)$. This idea originated in Strassen~\cite{strassen62} and is frequently used in the normal approximation regime.

Specifically, given any $\delta, \nu \in \bbR^+$, we define
\begin{align}
\cS_1(\delta, \nu) & \eqdef  \left\{P \in \cP(\cX) \colon \min_{P^\ast \in \cP^\ast_W} || P - P^\ast||_2 \leq \delta \textnormal{ and } V(P,W) \geq \nu \right\}, \label{eq:S1}\\
\cS_2(\delta, \nu) & \eqdef  \left\{P \in \cP(\cX) \colon \min_{P^\ast \in \cP^\ast_W} || P - P^\ast||_2 \leq \delta \textnormal{ and } V(P,W) < \nu \right\}, \label{eq:S2}\\
\cS_3(\delta) & \eqdef \left\{P \in \cP(\cX) \colon \min_{P^\ast \in \cP^\ast_W} || P - P^\ast||_2 > \delta  \right\}, \label{eq:S3}
\end{align}
where $\cP^\ast_W \eqdef \{ P \in \cP(\cX) \colon \mI(P;W) = C(W) \}$.

\begin{lemma}
Fix some $W \in \cP(\cY|\cX)$ with $C(W) >0$, $\delta \in \bbR^+$ and $\epsilon \in (0,1)$. Consider a sequence of constant composition $(N,R_N)$ codes $\{ (f_N, \varphi_N)\}_{N \geq 1}$ with the common composition $Q_N \in \cS_3(\delta)$ and
\begin{equation}
R_N \eqdef C(W) + \sqrt{\frac{V_\epsilon(W)}{N}} \Phi^{-1}(\epsilon). 	
\end{equation}
Then,
\begin{equation}
\bar{\mP}_{\me}(f_N, \varphi_N) > \epsilon,	
\end{equation}
for some $N_\mo(W, \epsilon, \delta) \in \bbZ^+$ and for all $N \geq N_{\mo}(W,\epsilon,\delta)$.
\label{lem:asym-lem1}
\end{lemma}
\begin{IEEEproof}
Define
\begin{equation}
\bbR^+ \ni \gamma(\delta) \eqdef C(W) - \sup_{Q \in \cS_3(\delta)}\mI(Q;W),
\end{equation}
Since $I(\cdot, W)$ is continuous over $\cP(\cX)$, $\gamma(\delta)$ is a well-defined and positive real number.
For any message $m$, let
\begin{equation}
G_N(m) \eqdef \left\{ \by^N : \frac{1}{N}\sum_{i=1}^N \ln \frac{W(y_i|x_i(m))}{q_{Q_N}(y_i)} > \mI(Q_N;W) + \frac{\gamma(\delta)}{2}\right\}.
\end{equation}
Define
\begin{equation}
\sigma^2_{\max} \eqdef \max_{P \in \cP(\cX)}V(P,W) \in \bbR^+.
\end{equation}
Since $V(\cdot, W)$ is continuous over the compact set $\cP(\cX)$ (e.g., \cite[Lemma~62]{polyanskiy10}), $\sigma^2_{\max}$ is a well-defined and positive real number.

The following arguments are essentially the ones used in \cite[Appendix~B]{altug12}, which we outline here for completeness. First,
\begin{equation}
\bar{\mP}_{\me}(f_N, \varphi_N) = 1 - \frac{1}{|\cM_N|}\sum_{m \in \cM_N}\sum_{\by^N \in \cA_m \cap G_N(m)}W(\by^N|\bx^N(m)) - \frac{1}{|\cM_N|}\sum_{m \in \cM_N}\sum_{\by^N \in \cA_m \cap G_N^c(m)}W(\by^N|\bx^N(m)). \label{eq:asym-lem1-pf-0}
\end{equation}
Since $q_{Q_N}$ is a probability measure on $\cY^N$ and the decoding regions are disjoint, one can verify that
\begin{equation}
\frac{1}{|\cM_N|}\sum_{m \in \cM_N}\sum_{\by^N \in \cA_m \cap G_N^c(m)}W(\by^N|\bx^N(m)) \leq \exp\left\{-N \left[ \frac{\gamma(\delta)}{2} + \sqrt{\frac{V_\epsilon(W)}{N}}\Phi^{-1}(\epsilon) \right]\right\}. \label{eq:asym-lem1-pf-1}
\end{equation}
Moreover, via an application of Chebyshev's inequality, it is easy to verify that
\begin{align}
\frac{1}{|\cM_N|}\sum_{m \in \cM_N}\sum_{\by^N \in \cA_m \cap G_N(m)}W(\by^N|\bx^N(m)) & \leq  \frac{N\cdot V(Q;W)}{\frac{(N\gamma(\delta))^2}{4}} \\
& \leq \frac{4 \sigma^2_{\max}}{N \gamma(\delta)^2}. \label{eq:asym-lem1-pf-2}
\end{align}
By substituting \eqref{eq:asym-lem1-pf-1} and \eqref{eq:asym-lem1-pf-2} into \eqref{eq:asym-lem1-pf-0} and choosing $N_\mo(W, \epsilon, \delta) \in \bbZ^+$ such that for all $N \geq N_\mo(W, \epsilon, \delta)$,
\begin{equation}
\bar{\mP}_{\me}(f_N, \varphi_N) \ge
1 - \exp\left\{-N\left[ \frac{\gamma(\delta)}{2} + \sqrt{\frac{V_\epsilon(W)}{N}}\Phi^{-1}(\epsilon)\right]\right\}- \frac{4 \sigma^2_{\max}}{N \gamma(\delta)^2},
\end{equation}
which tends to one as $n \rightarrow \infty$. This concludes the proof.
\end{IEEEproof}

\begin{lemma}
Fix some $\epsilon \in (\frac{1}{2},1)$, $W \in \cP(\cY|\cX)$ with $V_\epsilon(W) >0$, and $a \in \bbR^+$ with $a>\frac{2}{1-\epsilon}$. Consider an $(N,R_N)$ constant composition code $(f,\varphi)$ with
\begin{equation}
R_N = C(W) + \sqrt{\frac{V_\epsilon(W)}{N}}\Phi^{-1}(\epsilon) - \frac{1}{N}\ln\left( 1 - \epsilon - \frac{2}{a}\right),	
\end{equation}
and the common composition $Q$ satisfying
\begin{equation}
V(Q,W) < \frac{1}{a}V_{\epsilon}(W)\left[\Phi^{-1}(\epsilon)\right]^2. 	
\end{equation}
Then,
\begin{equation}
\bar{\mP}_{\me}(f, \varphi) > \epsilon. 	
\end{equation}
\label{lem:asym-lem2}
\end{lemma}
\vspace{-0.5cm}
\begin{IEEEproof}
Via arguments similar to the ones given in the proof of Lemma~\ref{lem:asym-lem1}, one can verify that
\begin{align}
\bar{\mP}_{\me}(f, \varphi) & \geq 1 - \left( 1 - \epsilon - \frac{2}{a} \right) - \frac{N\cdot V(Q,W)}{\left[ N[C(W) - \mI(Q;W)] + \sqrt{N \cdot V_\epsilon(W)}\Phi^{-1}(\epsilon)\right]^2}  \\
& \geq \epsilon + \frac{1}{a}\\
& > \epsilon.
\end{align}
\end{IEEEproof}

For any $Q \in \cP(\cX)$, define
\begin{align}
\label{eq:Udef}
U(Q,W) & \eqdef \sum_{(x,y) \in \cX \times \cY}Q(x)W(y|x)  \left[ \ln \frac{W(y|x)}{q_Q(y)} - \mI(Q;W) \right]^2, \\
m_3(Q,W) & \eqdef \sum_{x \in \cX}Q(x)\mE_{W(\cdot|x)}\left[ \left| \ln \frac{W(Y|x)}{q_Q(Y)} - \mE_{W(\cdot|x)}\left[ \ln \frac{W(Y|x)}{q_Q(Y)}\right] \right|^3\right].
\end{align}
Choose $\delta >0$ such that\footnote{As usual, without loss of generality, we assume that $W$ has no all-zero columns.}
\begin{equation}
 \supp(q_Q) = \cY,  \textnormal{ for all } Q \in \cP(\cX) \backslash \cS_3(\delta). \label{eq:asym-delta}
\end{equation}
Such a choice is possible due to the evident continuity of $\alpha_{y}(\cdot)$ for any $y \in \cY$ and the fact that the unique capacity achieving output distribution has full support, as noted before. The following has been shown by Polyanskiy \emph{et al.} \cite[Lemma~46]{polyanskiy10}
\begin{align}
\tilde{m}_3(Q,W) & \eqdef \sum_{(x,y) \in \cX \times \cY}Q(x)W(y|x)\left| \ln \frac{W(Y|X)}{q_Q(Y)} - \mI(Q;W) \right|^3 \\
& \leq \left(\frac{3}{e}\left( |\cX|^{1/3} + |\cY|^{1/3}\right) + \ln \min\{ |\cX|, |\cY| \}\right)^3 \label{eq:asym-alpha.0}\\
& =: \kappa(W) \in \bbR^+. \label{eq:asym-alpha.1}
\end{align}

Fix some $\nu \in \bbR^+$ and $\epsilon \in (0,1)$. Assume $\cS_1(\delta, \nu) \neq \emptyset$ and define
\begin{equation}
K(W, \epsilon, \delta, \nu) \eqdef \frac{2}{\phi(\Phi^{-1}(\epsilon))}\left[ \max_{P \in \cS_1(\delta, \nu)}\frac{m_3(P,W)}{V(P,W)} + \left( \frac{1}{\sqrt{2\pi}} + \frac{\kappa(W)}{\nu}\right) \right]\in \bbR^+. \label{eq:asym-K}
\end{equation}
Since $m_3(\cdot,W)$ and $V(\cdot, W)$ are continuous over $\cP(\cX)$ (e.g., \cite[Lemma~62]{polyanskiy10}), $K(W, \epsilon, \delta, \nu)$ is a well-defined and positive real number.

\begin{lemma}
Fix an asymmetric and singular $W \in \cP(\cY|\cX)$, $\epsilon \in (0, 1)$ and $\nu \in \bbR^+$. Choose $\delta \in \bbR^+$ such that \eqref{eq:asym-delta} holds. For some $\tilde{N}_\mo(W, \epsilon, \delta, \nu) \in \bbZ^+$ and any $N \geq \tilde{N}_\mo(W, \epsilon, \delta, \nu)$, consider an $(N,R_N)$ constant composition code $(f, \varphi)$ with common composition $Q \in \cS_1(\delta, \nu)$ and
\begin{equation}
R_N = \mI(Q;W) + \sqrt{\frac{V(Q,W)}{N}} \Phi^{-1}(\epsilon) + \frac{1}{N}K(W, \epsilon, \delta, \nu).	
\end{equation}
Then $\bar{\mP}_{\me}(f, \varphi) >\epsilon$.
\label{lem:asym-lem3}
\end{lemma}

\begin{IEEEproof}
Assume $\cS_1(\delta, \nu) \neq \emptyset$, because otherwise the claim is void. The proof is similar to the proof of Theorem~\ref{thrm:sym}. Let $\tilde{N}_\mo(W, \epsilon, \delta, \nu) \in \bbZ^+$ be such that for all $N \geq \tilde{N}_\mo(W, \epsilon, \delta, \nu)$,
\begin{equation}
\sqrt{N} > \frac{2 K(W, \epsilon, \delta, \nu)}{\phi(\Phi^{-1}(\epsilon))\sqrt{\nu}}.
\end{equation}
In light of \eqref{eq:asym-K}, the existence of such a choice is evident.

Consider any $(N,R_N)$ constant composition code, say $(f, \varphi)$, with the common composition $Q$. Assume $Q$ and $R_N$ are as in the statement of the lemma. Consider any $\bx^N \in \cX^N$ and define
\begin{equation}
M_{\bx^N}(\lambda) \eqdef \mE_{W(\cdot|\bx^N)}\left[ \me^{\lambda \ln \frac{W(\bY^N|\bx^N)}{q_{P_{\bx^N}}(\bY^N)}}\right], \, \forall \, \lambda \in \bbR.
\end{equation}
We claim that for any $\bx^N, \bz^N \in \cX^N$ with $P_{\bx^N} = P_{\bz^N}$, we have
\begin{equation}
M_{\bx^N}(\lambda) = M_{\bz^N}(\lambda), \, \forall \, \lambda \in \bbR. \label{eq:asym-lem3-pf0}
\end{equation}
To see this, we simply note that
\begin{align}
M_{\bx^N}(\lambda) & = \sum_{\by^N : W(\by^N | \bx^N) >0}\me^{N \sum_{y}P_{\by^N}(y) \ln \xi_y} \me^{-\lambda N \sum_{y}P_{\by^N}(y) \ln \alpha_y(P_{\bx^N})} \\
& = \sum_{P \in \cP_N(\cY)}\me^{N \sum_{y}P(y) \ln \xi_y} \me^{-\lambda N \sum_{y}P(y) \ln \alpha_y(P_{\bx^N})}|\{ \by^N : P_{\by^N} = P \textnormal{ and } W(\by^N |\bx^N) >0\}| \\
& = \sum_{P \in \cP_N(\cY)}\me^{N \sum_{y}P(y) \ln \xi_y} \me^{-\lambda N \sum_{y}P(y) \ln \alpha_y(P_{\bz^N})}|\{ \by^N : P_{\by^N} = P \textnormal{ and } W(\by^N |\bz^N) >0\}| \label{eq:asym-lem3-pf1}\\
& = M_{\bz^N}(\lambda), \label{eq:asym-lem3-pf1.1}
\end{align}
where \eqref{eq:asym-lem3-pf1} follows from the fact that $P_{\bx^N} = P_{\bz^N}$. Equation~\eqref{eq:asym-lem3-pf0}, along with the uniqueness theorem for the moment generating function (e.g., \cite[Ex.~26.7]{billingsley95}), and the fact that $q_Q$ is of full support, enables us to invoke Lemma~\ref{lem:multipurp} to deduce that
\begin{equation}
\bar{\mP}_{\me}(f, \varphi) \geq W(\cS_{R_N}(Q) | \bz^N) - \sum_{\by^N \in \cS_{R_N}(Q)}q_Q(\by^N) \exp\left\{-N\left[ R_N - \frac{1}{N}\sum_{i=1}^N \ln \frac{1}{\alpha_{y_i}(Q)}\right]\right\}, \label{eq:asym-lem3-pf2}
\end{equation}
for a given $\bz^N \in \cX^N$ with $P_{\bz^N} = Q$. Due to the singularity of $W$,
\begin{align}
W(\cS_{R_N}(Q) | \bz^N) & = \sum_{\by^N}W(\by^N | \bz^N)\b1\left\{ \frac{1}{N}\sum_{i=1}^N \ln \frac{W(y_i|z_i)}{q_Q(y_i)} \leq R_N \right\} \\
& \geq \epsilon - \frac{m_3(Q,W)}{\sqrt{N}V(Q,W)^{3/2}} + \frac{K(W, \epsilon, \delta, \nu) \phi(\Phi^{-1}(\epsilon))}{\sqrt{N \cdot V(Q,W)}}\left( 1 - \frac{K(W, \epsilon, \delta, \nu) }{2\sqrt{N \cdot V(Q,W)}\phi(\Phi^{-1}(\epsilon))}\right), \label{eq:asym-lem3-pf3}
\end{align}
where the proof of \eqref{eq:asym-lem3-pf3} is similar to that of \eqref{eq:sym-pf-final5} and omitted for brevity.

Further, define
\begin{align}
P_{XY}(x,y) & \eqdef Q(x)W(y|x), \\
P_{\bX^N\bY^N}(\bx^n, \by^n) & \eqdef \prod_{i=1}^N P_{XY}(x_i, y_i).	
\end{align}
Evidently,
    \begin{align}
        &     \sum_{\by^N \in \cS_{R_N}(Q)}q_Q(\by^N) \exp\left\{-N\left[ R_N - \frac{1}{N}\sum_{i=1}^N \ln \frac{1}{\alpha_{y_i}(Q)}\right]\right\}  \\
           & = \sum_{(\bx^N, \by^N)}P_{\bX^N\bY^N}(\bx^N, \by^N)\b1\left\{\frac{1}{N}\sum_{i=1}^N \ln \frac{W(y_i|x_i)}{q_Q(y_i)} \leq R_N \right\} \nonumber \\
& \quad \times  \exp\left\{-N\left[R_N - \frac{1}{N}\sum_{i=1}^N \ln \frac{W(y_i|x_i)}{q_Q(y_i)}\right]\right\} \\
     & \leq \frac{1}{\sqrt{2\pi N \cdot U(Q,W)}} + \frac{\tilde{m}_3(Q,W)}{\sqrt{N}U(Q,W)^{3/2}} \label{eq:asym-lem3-pf4}\\
     & \leq \frac{1}{\sqrt{N \cdot V(Q,W)}}\left( \frac{1}{\sqrt{2\pi}} + \frac{\kappa(W)}{V(Q,W)}\right), \label{eq:asym-lem3-pf5}
    \end{align}
where $U(Q,W)$ is defined in \eqref{eq:Udef}, and
\eqref{eq:asym-lem3-pf4} follows from Lemma~\ref{lem:lem2}, whose application is ensured by the fact that $U(Q,W) \geq V(Q,W)$ (e.g., \cite[Lemma~62]{polyanskiy10}), which, along with \eqref{eq:asym-alpha.1}, also implies \eqref{eq:asym-lem3-pf5}.

By substituting \eqref{eq:asym-lem3-pf3} and \eqref{eq:asym-lem3-pf5} into \eqref{eq:asym-lem3-pf2}, along with the definitions of $K(W,\epsilon, \delta, \nu)$ and $n_\mo(W, \epsilon, \delta, \nu)$, one can verify that
\begin{align}
\bar{\mP}_{\me}(f,\varphi) & > \epsilon + \frac{1}{\sqrt{N \cdot V(Q,W)}}\left( \max_{P \in \cS_1(\delta,\nu)} \frac{m_3(P,W)}{V(P,W)} - \frac{m_3(Q,W)}{V(Q,W)}\right) \\
& \geq \epsilon,
\end{align}
which, in turn, implies the assertion.
\end{IEEEproof}

In order to prove the first assertion of the theorem, i.e., \eqref{eq:thrm-asym-1}, fix some $\epsilon \in (0,\frac{1}{2})$ and assume $V_\epsilon(W) >0$, because otherwise \cite[Proposition~9]{tomamichel-tan13} implies \eqref{eq:thrm-asym-1}. Fix some $\delta >0$ such that \eqref{eq:asym-delta} holds and $\cS_2\left(\delta, \frac{V_\epsilon(W)}{2}\right) = \emptyset$. Such a choice is possible since $V(\cdot,W)$ is continuous over $\cP(\cX)$, as noted before. For any $P \in \cP(\cX)$, let
\begin{equation}
P^{\ast}(P) \eqdef \arg \min_{Q \in \cP^\ast_W}|| Q - P||_2. 	
\end{equation}
Fix some $\beta_1, \beta_2 \in \bbR^+$ such that
\begin{align}
\mI(P;W) & \leq C(W) - \beta_1||P - P^\ast(P) ||_2^2, \label{eq:asym-beta.0}\\
|\sqrt{V(P,W)} - \sqrt{V(P^\ast(P),W)}| &\leq \beta_2 ||P - P^\ast(P) ||_2,
\label{eq:asym-beta.1}
\end{align}
for any $P \in \cS_1\left(\delta, \frac{V_\epsilon(W)}{2}\right)$, whose existence is ensured by \cite[Lemma~7]{tomamichel-tan13}. In light of \eqref{eq:asym-beta.0} and \eqref{eq:asym-beta.1}, for all $P \in \cS_1\left(\delta, \frac{V_\epsilon(W)}{2}\right)$ and for any $N \in \bbZ^+$,
\begin{align}
N\mI(P;W) + \sqrt{N \cdot V(P,W)}\Phi^{-1}(\epsilon) & \leq N\cdot C(W) + \sqrt{N \cdot V_\epsilon(W)}\Phi^{-1}(\epsilon) \nonumber \\
& \quad  -\beta_1  N ||P - P^\ast(P) ||_2^2 + \beta_2 |\Phi^{-1}(\epsilon)|\sqrt{N}||P - P^\ast(P) ||_2  \\
& \leq N\cdot C(W) + \sqrt{N \cdot V_\epsilon(W)}\Phi^{-1}(\epsilon) + \frac{1}{4 \beta_1}\left(\beta_2 |\Phi^{-1}(\epsilon)|\right)^2, \label{eq:asym-final1}
\end{align}
where \eqref{eq:asym-final1} follows from elementary calculus. Consider any $N \in \bbZ^+$ such that
\begin{equation}
N \geq \max\left\{N_\mo(W, \epsilon, \delta), \tilde{N}_\mo(W, \epsilon, \delta, \tfrac{V_\epsilon(W)}{2})\right\},
\end{equation}
where $N_\mo$ and $\tilde{N}_\mo$ are given in Lemmas~\ref{lem:asym-lem1} and \ref{lem:asym-lem3}, respectively. Define
\begin{equation}
R_N \eqdef C(W) + \sqrt{\frac{V_\epsilon(W)}{N}}\Phi^{-1}(\epsilon) + \frac{1}{N}\left( \frac{1}{4 \beta_1}\left(\beta_2 |\Phi^{-1}(\epsilon)|\right)^2 + K(W, \epsilon, \delta, \tfrac{V_\epsilon(W)}{2}) \right),
\end{equation}
and consider any $(N,R_N)$ constant composition code $(f, \varphi)$ with the common composition $Q$. Now, if $Q \in \cS_3(\delta)$, then Lemma~\ref{lem:asym-lem1} implies that $\bar{\mP}_{\me}(f, \varphi) >\epsilon$. Similarly, if $Q \in \cS_1\left(\delta, \frac{V_\epsilon(W)}{2}\right) $, then Lemma~\ref{lem:asym-lem3} and \eqref{eq:asym-final1} imply that $\bar{\mP}_{\me}(f, \varphi) >\epsilon$. Since the code is arbitrary, we conclude that \eqref{eq:thrm-asym-1} holds.

In order to prove the second assertion of the theorem, i.e., \eqref{eq:thrm-asym-2}, fix some $\epsilon \in (\frac{1}{2}, 1)$ and $\delta >0$ such that \eqref{eq:asym-delta} holds. Choose some $a \in \bbR^+$ that satisfies $a > \frac{2}{1-\epsilon}$ and $\nu \in \bbR^+$ such that $\nu \leq \frac{1}{a}V_\epsilon(W)\left[\Phi^{-1}(\epsilon)\right]^2$. Similar to \eqref{eq:asym-beta.0} and \eqref{eq:asym-beta.1}, choose $\beta_1, \beta_2 \in \bbR^+$ such that
\begin{align}
\mI(P;W) & \leq C(W) - \beta_1||P - P^\ast(P) ||_2^2, \label{eq:asym-beta-1.0}\\
 |\sqrt{V(P,W)} - \sqrt{V(P^\ast(P),W)}| & \leq \beta_2 ||P - P^\ast(P) ||_2,
\label{eq:asym-beta-1.1}
\end{align}
for any $P \in \cS_1\left(\delta, \nu \right)$. From \eqref{eq:asym-beta-1.0} and \eqref{eq:asym-beta-1.1}, similar to \eqref{eq:asym-final1}, we deduce that for all $P \in \cS_1(\delta, \nu)$ and $N \in \bbZ^+$,
\begin{equation}
N \cdot \mI(P;W) + \sqrt{N \cdot V(P,W)}\Phi^{-1}(\epsilon) \leq  N\cdot C(W) + \sqrt{N \cdot V_\epsilon(W)}\Phi^{-1}(\epsilon) + \frac{1}{4 \beta_1}\left(\beta_2 \Phi^{-1}(\epsilon)\right)^2. \label{eq:asym-final2}
\end{equation}
Consider any $N \in \bbZ^+$ such that
\begin{equation}
N \geq \max\{  N_\mo(W, \epsilon, \delta),   \tilde{N}_\mo(W, \epsilon, \delta, \nu) \}, 	
\end{equation}
where $N_\mo$ and $\tilde{N}_\mo$ are as given in Lemmas~\ref{lem:asym-lem1} and \ref{lem:asym-lem3}, respectively. Consider any $(N,R_N)$ constant composition code $(f, \varphi)$ with the common composition $Q$ and define
\begin{equation}
R_N \eqdef C(W) + \sqrt{\frac{V_\epsilon(W)}{N}}\Phi^{-1}(\epsilon) + \frac{1}{N}\left( \frac{1}{4 \beta_1}\left(\beta_2 \Phi^{-1}(\epsilon)\right)^2 + K(W, \epsilon, \delta, \nu) - \ln \left(1-\epsilon- \frac{2}{a}\right) \right).
\end{equation}
If $Q \in \cS_3(\delta)$, then $\bar{\mP}_\me(f, \varphi) > \epsilon$ due to Lemma~\ref{lem:asym-lem1}. If $Q \in \cS_2(\delta, \nu)$, then $\bar{\mP}_\me(f, \varphi) > \epsilon$ because of Lemma~\ref{lem:asym-lem2}. Finally, if $Q \in \cS_1(\delta, \nu)$, then Lemma~\ref{lem:asym-lem3}, along with \eqref{eq:asym-final2}, implies that $\bar{\mP}_\me(f, \varphi) > \epsilon$. Since the code is arbitrary, we conclude that \eqref{eq:thrm-asym-2} holds.
\hfill \IEEEQED

\section{Discussion}
\label{sec:discussion}
\subsection{Relation to the minimax converse}
\label{ssec:minmax}
In the absence of feedback, one can interpret the proof of Theorem~\ref{thrm:sym} in terms of the minimax converse (e.g., \cite[Theorem~1]{polyanskiy13}), which we illustrate next. To this end, we fix a symmetric and singular $W \in \cP(\cY|\cX)$ and note that \cite[Eq.~(9) and (11)]{polyanskiy13} imply that for any $N \in \bbZ^+$ and $\epsilon \in (0,1)$,
\begin{equation}
\min_{P_{\bX^N}} \max_{Q_{\bY^N}} \beta_{1-\epsilon}(P_{\bX^N \bY^N}, P_{\bX^N} \times Q_{\bY^N}) \leq \frac{1}{M^\ast(N, \epsilon)}, \label{eq:conc-minmax-1}
\end{equation}
where
\begin{align}
P_{\bX^N \bY^N}(\bx^N, \by^N) & \eqdef P_{\bX^N}(\bx^N)W(\by^N|\bx^N), \\
(P_{\bX^N} \times Q_{\bY^N})(\bx^N, \by^N) & \eqdef P_{\bX^N}(\bx^N)Q_{\bY^N}(\by^N),	
\end{align}
and $\beta_{1-\epsilon}(P_{\bX^N \bY^N}, P_{\bX^N} \times Q_{\bY^N})$ denotes the minimum probability of error under $P_{\bX^N} \times Q_{\bY^N}$, subject to the constraint that the error probability under hypothesis $P_{\bX^N \bY^N}$ does not exceed $\epsilon$. Due to \cite[Theorem~21]{polyanskiy13}, the minimum on the left side of \eqref{eq:conc-minmax-1} is attained by $U_{\cX^N}$. Consider some $N \in \bbZ^+$ such that \eqref{eq:sym-pf-final1} holds and let $R$ be as in \eqref{eq:sym-pf-final2}. With these choices, we define\footnote{The non-product distribution in \eqref{eq:conc-minimax-2} is inspired by \cite[Eq.~(168)]{polyanskiy13}. In particular, if $W$ is BEC then \eqref{eq:conc-minimax-2} reduces to \cite[Eq.~(168)]{polyanskiy13}.}
\begin{equation}
Q^\ast_{\bY^N}(\by^N) \eqdef \frac{\me^{N \sum_{y}P_{\by^N}(y) \ln \xi_y } \b1\left\{ \by^N \in \cS(R) \right\}}{\sum_{\bb^N}\me^{N \sum_{b}P_{\bb^N}(b) \ln \xi_b } \b1\left\{ \bb^N \in \cS(R) \right\}},
\label{eq:conc-minimax-2}
\end{equation}
where $\xi_y$ and $\cS(R)$ are as defined before. Evidently,
\begin{equation}
Q^\ast_{\bY^N} \in \cP(\cY^N). 	
\end{equation}
With a slight abuse of notation, let $\beta_{1 - \epsilon}(U_{\cX^N}, Q^\ast_{\bY^N})$ denote the value of the cost function of the optimization problem in \eqref{eq:conc-minmax-1} when $P_{\bX^N} = U_{\cX^N}$ and $Q_{\bY^N} = Q^\ast_{\bY^N}$. Evidently,
\begin{equation}
M^\ast(N, \epsilon) \leq \frac{1}{\beta_{1 - \epsilon}(U_{\cX^N}, Q^\ast_{\bY^N})}.
\label{eq:conc-minimax-3}
\end{equation}
From the Neyman-Pearson lemma (e.g., \cite{neyman-pearson33}), the right side of \eqref{eq:conc-minimax-3} is attained by a randomized threshold test with the randomization parameter $\tau \in (0,1)$ satisfying 
\begin{align}
\tau W(\cS(R) | \bx^N_\mo) & = \epsilon, \label{eq:conc-minimax-4.0}\\
\beta_{1 - \epsilon}(U_{\cX^N}, Q^\ast_{\bY^N}) & = \frac{(1-\tau)W(\cS(R) | \bx^N_\mo)}{\me^{NR} \sum_{\by^N \in \cS(R)}q(\by^N)\exp\left\{-N\left[R - \frac{1}{N}\sum_{i=1}^N \ln \frac{1}{\alpha_{y_i}}\right]\right\}} \label{eq:conc-minimax-4.1}.
\end{align}
Equations~\eqref{eq:conc-minimax-4.0} and \eqref{eq:conc-minimax-4.1} can be verified via elementary algebra by noticing that $W$ is singular and symmetric. We omit the details for brevity. Finally,  \eqref{eq:sym-pf-final4} and \eqref{eq:sym-pf-final5}, along with \eqref{eq:sym-pf-final1} and \eqref{eq:sym-pf-final2}, imply that
\begin{equation}
W(\cS(R)|\bx^N_\mo) - \sum_{\by^N \in \cS(R)}q(\by^N)\exp\left\{-N\left[R - \frac{1}{N}\sum_{i=1}^N \ln \frac{1}{\alpha_{y_i}}\right]\right\} > \epsilon. \label{eq:conc-minimax-5}
\end{equation}
Equations~\eqref{eq:conc-minimax-3}--\eqref{eq:conc-minimax-5} imply that $M^\ast(N, \epsilon) < e^{NR}$, which, in turn, implies Theorem~\ref{thrm:sym} in the absence of feedback.

The above interpretation of the arguments leading to \eqref{eq:conc-minimax-5} yield a more streamlined alternative to the one in the main text, at least for the case of no feedback. We have provided the latter because it allows for feedback and because it gives a unified method for proving converse results in the fixed-rate and fixed-error-probability regimes.

\subsection{On dropping the constant composition assumption}
\label{ssec:CCC}
As noted before, Theorem~\ref{thrm:asym} gives an $O(1)$ upper bound on the third-order term of the normal approximation for asymmetric and singular DMCs only if we consider constant composition codes. Although this restriction is undesirable, it is quite common in converse results. Indeed, the usual proof of the converse statement of \eqref{eq:intro-1} involves first showing it for constant composition codes, and then arguing that this restriction at most results in an extra $O(\ln N)$ term.

        Tomamichel and Tan~\cite{tomamichel-tan13} have showed an $\ln \sqrt{N} $ upper bound on the third-order term in general by eliminating the constant composition code restriction in the first step. This result, coupled with the existing results in the literature, gives the third-order term for a broad class of channels, which includes positive channels with positive capacity but does not include asymmetric and singular channels. The method of \cite{tomamichel-tan13} is based on relating the channel coding problem to a binary hypothesis test by using an auxiliary output distribution, which is in the same vein as the so-called meta-converse of Polyanskiy \emph{et al.} (e.g., \cite[Section~III.E and III.F]{polyanskiy10}). As opposed to the classical applications of this idea, which use a product auxiliary output distribution and result in the aforementioned two-step procedure, the authors of \cite{tomamichel-tan13} uses an appropriately chosen non-product output distribution to dispense with the constant composition step. However, their non-product distribution is different from the one used in the previous subsection. Investigating how to combine the analysis of \cite{tomamichel-tan13} and the viewpoint in Section~\ref{ssec:minmax} to drop the constant composition assumption in Theorem~\ref{thrm:asym} is a worthy direction for future research.

\subsection{Limitation in the error exponents regime}
\label{ssec:EA}
 One might conjecture that by following the same program used to prove Theorem~\ref{thrm:asym}, one could prove the following lower bound for asymmetric and singular channels
\begin{equation}
\liminf_{N \rightarrow \infty}\frac{\bar{\mP}_{\me, \textnormal{c}}(N,R)}{\frac{1}{\sqrt{N}}\me^{-N \mE_{\mSP}(R)}}
 \geq K(R,W), \label{eq:conc-EA}
\end{equation}
where $K(R,W)$ is a positive constant that depends on $R$ and $W$. However, a proof of \eqref{eq:conc-EA} seems to be more involved than its counterpart in the normal approximation regime, i.e., Theorem~\ref{thrm:asym}. The main technical difficulty is proving the continuity properties of $\mE_{\mSP}(R,\cdot)$ that are required to distinguish between the ``good types'', for which $\mE_{\mSP}(R,Q) \approx \mE_{\mSP}(R)$ and hence one can use a result like Lemma~\ref{lem:asym-lem3} to deduce an $\Omega(\frac{1}{\sqrt{N}})$ sub-exponential term directly, and the ``bad types'', for which $\mE_{\mSP}(R,Q)$ is bounded away from $\mE_{\mSP}(R)$ and hence one can utilize this inferiority of the exponent to deduce an $\Omega(\frac{1}{\sqrt{N}})$ sub-exponential term. Indeed, justifications of these continuity properties appear to be quite intricate. For 
an analogous upper bound, see Honda~\cite{Honda:RC:ISIT15,Honda:RC:ISIT18}.

\appendices

\section{Proof of Proposition~\ref{lem:SP-optimizers-new}}
\label{app-SP-opt}
\begin{itemize}
\item[(i)] Thanks to the symmetry of the channel, $\tilde{\mE}_{\mSP}(R) = \tilde{\mE}_{\mSP}(R, U_\cX)$ (e.g., \cite[p.~145]{gallager68}). Moreover, due to the facts that $\mE_{\mSP}(R) = \tilde{\mE}_{\mSP}(R)$ and $\mE_{\mSP}(R,P) \geq \tilde{\mE}_{\mSP}(R,P)$ for all $P \in \cP(\cX)$, which have been noted before, we conclude that $\mE_{\mSP}(R) = \mE_{\mSP}(R, U_\cX)$.
\item[(ii)] Fix any $\rho \in \bbR_+$ and consider the following convex program
\begin{equation}
\min_{Q \in \cP(\cX)}\sum_{y \in \cY}\left( \sum_{x \in \cX}Q(x)W(y|x)^{1/(1+\rho)}\right)^{1+\rho},
\label{eq:SP-optimizers-pf1}
\end{equation}
whose convexity is verified in \cite[Theorem~5.6.5]{gallager68}. Next, we recall the necessary and sufficient conditions for any $Q \in \cP(\cX)$ to attain the minimum in \eqref{eq:SP-optimizers-pf1}, due to \cite[Theorem~5.6.5]{gallager68},
\begin{equation}
\forall x \in \cX, \, \sum_{y \in \cY}W(y|x)^{1/(1+\rho)} \left(\sum_{z \in \cX} Q(z)W(y|z)^{1/(1+\rho)} \right)^{\rho} \geq \sum_{y \in \cY}\left( \sum_{z \in \cX} Q(z)W(y|z)^{1/(1+\rho)}\right)^{1+\rho},
\label{eq:lem-SP-optimizers-pf3}
\end{equation}
with equality if $Q(x) >0$. Thanks to the symmetry of the channel, $U_\cX$ is an optimizer of \eqref{eq:SP-optimizers-pf1} (e.g., \cite[p.~145]{gallager68}) and hence \eqref{eq:lem-SP-optimizers-pf3} implies \eqref{eq:SP-optimizers-1}.
\item[(iii)] We first note the following, which is an easy consequence of elementary convex optimization arguments (e.g., \cite[Ex.~2.5.23]{csiszar-korner81})
\begin{equation}
\mE_{\mSP}(R, U_\cX) = \max_{\rho \, \geq \, 0} \min_{q \in \cP(\cY)} \left\{ -\rho R - (1+\rho)\sum_{x \in \cX} U_\cX(x) \ln \sum_{y \in \cY} W(y|x)^{1/(1+\rho)}q(y)^{\rho/(1+\rho)}\right\}. \label{eq:lem-SP-optimizers-pf4}
\end{equation}
Due to \cite[Propositions~1 and 2]{altug12a}, \eqref{eq:lem-SP-optimizers-pf4} has a unique saddle-point. Further, \cite[Proposition~3]{altug12a} ensures that $\rho_R $ is the $\bbR_+$ component of this saddle-point. Owing to the properties of the saddle-points (e.g., \cite[Lemma~36.2]{rockafellar70}) $\rho_R$ attains the maximum in \eqref{eq:lem-SP-optimizers-pf4}, and the fact that $\mE_{\mSP}(R) = \mE_{\mSP}(R,U_\cX) >0$ ensures its positivity. Hence,
\begin{align}
\mE_{\mSP}(R, U_\cX) & = \min_{q \in \cP(\cY)} \left\{ -\rho_R R - (1+\rho_R)\sum_{x \in \cX} U_\cX(x) \ln \sum_{y \in \cY} W(y|x)^{1/(1+\rho_R)}q(y)^{\rho_R/(1+\rho_R)}\right\} \label{eq:lem-SP-optimizers-pf5.0} \\
& \leq  -\rho_R R - (1+\rho_R)\sum_{x \in \cX} U_\cX(x) \ln \sum_{y \in \cY} W(y|x)^{1/(1+\rho_R)}q_R(y)^{\rho_R/(1+\rho_R)}  \\
& = -\rho_R R + \mE_{\mo}(\rho_R, U_\cX) \label{eq:lem-SP-optimizers-pf5}\\
& \leq \tilde{\mE}_{\mSP}(R, U_\cX), \label{eq:lem-SP-optimizers-pf6}
\end{align}
where \eqref{eq:lem-SP-optimizers-pf5} follows from the second assertion of this proposition, i.e., \eqref{eq:SP-optimizers-1}, along with the definitions of $q_R$ and $\mE_{\mo}(\cdot, \cdot)$. In light of the first assertion of this proposition, i.e., \eqref{eq:SP-optimizers-item-i}, \eqref{eq:lem-SP-optimizers-pf6} implies that $\rho_R$ attains the maximum in the definition of $\tilde{\mE}_{\mSP}(R,U_\cX)$.
\item[(iv)] Equation~\eqref{eq:lem-SP-optimizers-pf6} and the first assertion of this proposition ensure that $q_R$ attains the minimum in \eqref{eq:lem-SP-optimizers-pf5.0}. Hence, by recalling the definition of a saddle-point (e.g., \cite[p.~380]{rockafellar70}), in order to conclude the proof, it suffices to show that $\rho_R$ attains the supremum in the following optimization problem:
\begin{equation}
\sup_{\rho \in \bbR_+}\left\{ -\rho R - (1+\rho)\sum_{x \in \cX}U_{\cX}(x)\ln \sum_{y \in \cY}W(y|x)^{1/(1+\rho)}q_R(y)^{\rho/(1+\rho)}\right\}. \label{eq:lem-SP-optimizers-pf7}
\end{equation}
 To this end, for any $\rho \in \bbR_+$, define
 \begin{align}
q_{\rho}(y) & \eqdef \frac{\left(\sum_{x \in \cX} U_{\cX}(x)W(y|x)^{1/(1+\rho)}\right)^{1+\rho}}{\sum_{b \in \cY}\left(\sum_{a \in \cX} U_{\cX}(a)W(b|a)^{1/(1+\rho)}\right)^{1+\rho}}, \label{eq:lem-SP-optimizers-pf8}\\
V_{\rho}(y|x) & \eqdef \frac{W(y|x)^{1/(1+\rho)}q_\rho(y)^{\rho/(1+\rho)}}{\sum_{b \in \cY}W(b|x)^{1/(1+\rho)}q_\rho(b)^{\rho/(1+\rho)}}. \label{eq:lem-SP-optimizers-pf9}
\end{align}
Recalling the definition of $q_R$, i.e., \eqref{eq:qR}, along with \eqref{eq:lem-SP-optimizers-pf8}, we notice that $q_R = q_{\rho_R}$. We proceed by noting that
\begin{align}
\sum_{x \in \cX}U_{\cX}(x)V_{\rho}(y|x) & = \sum_{x \in \cX}U_{\cX}(x)\frac{W(y|x)^{\frac{1}{1+\rho}}\left[ \sum_{z \in \cX}U_{\cX}(z)W(y|z)^{\frac{1}{1+\rho}}\right]^{\rho}}{\sum_{b \in \cY}W(b|x)^{\frac{1}{1+\rho}}\left[ \sum_{a \in \cX}U_{\cX}(a)W(b|a)^{\frac{1}{1+\rho}}\right]^{\rho}} \label{eq:lem-SP-optimizers-pf11.0}\\
& = \frac{\sum_{x \in \cX}U_{\cX}(x)W(y|x)^{\frac{1}{1+\rho}}\left[\sum_{z \in \cX}U_{\cX}(z)W(y|z)^{\frac{1}{1+\rho}} \right]^{\rho}}{\sum_{b \in \cY}\left[ \sum_{a \in \cX}U_{\cX}(a)W(b|a)^{\frac{1}{1+\rho}}\right]^{1+\rho}}\label{eq:lem-SP-optimizers-pf11.00}\\
& = q_\rho(y), \label{eq:lem-SP-optimizers-pf11}
\end{align}
where \eqref{eq:lem-SP-optimizers-pf11.0} follows by substituting \eqref{eq:lem-SP-optimizers-pf8} into \eqref{eq:lem-SP-optimizers-pf9}, \eqref{eq:lem-SP-optimizers-pf11.00} follows from \eqref{eq:SP-optimizers-1}, which is verified in item (ii) of this proposition, and \eqref{eq:lem-SP-optimizers-pf11} follows from the definition of $q_\rho$, i.e., \eqref{eq:lem-SP-optimizers-pf8}. Note that
\begin{equation}
\mI(U_{\cX};V_{\rho}) = \mD(V_{\rho}\|q_{\rho}|U_{\cX}), \label{eq:lem-SP-optimizers-pf11.1}
\end{equation}
which is a direct consequence of the non-negativity of the relative entropy, along with \eqref{eq:lem-SP-optimizers-pf11}.

Next, we note that for any $\rho \in \bbR_+$,
\begin{equation}
\mD(V_{\rho}\|W|U_{\cX}) + \rho \mI(U_{\cX};V_\rho) = -(1+\rho)\sum_{x \in \cX}U_{\cX}(x)\ln \sum_{y \in \cY}W(y|x)^{1/(1+\rho)}q_{\rho}(y)^{\rho/(1+\rho)}. \label{eq:lem-SP-optimizers-pf10}
\end{equation}
To see \eqref{eq:lem-SP-optimizers-pf10}, first observe that
\begin{equation}
\mD(V_{\rho}\|W|U_{\cX}) = \sum_{x \in \cX}U_{\cX}(x)\sum_{y \in \cY}V_{\rho}(y|x)\left\{ \frac{\rho}{(1+\rho)}\ln\frac{q_{\rho}(y)}{W(y|x)} - \ln \sum_{b \in \cY}W(b|x)^{1/(1+\rho)}q_\rho(b)^{\rho/(1+\rho)}\right\}, \label{eq:lem-SP-optimizers-pf10.1}
\end{equation}
which is a direct consequence of the definition of $V_{\rho}(y|x)$, i.e., \eqref{eq:lem-SP-optimizers-pf9}. Further, \eqref{eq:lem-SP-optimizers-pf9}, coupled with \eqref{eq:lem-SP-optimizers-pf11.1}, implies that
\begin{equation}
\rho \mI(U_{\cX};V_{\rho}) = \rho\left[\sum_{x \in \cX}U_{\cX}(x)\sum_{y \in \cY}V_{\rho}(y|x)\left\{ \frac{1}{(1+\rho)}\ln \frac{W(y|x)}{q_\rho(y)} -  \ln \sum_{b \in \cY}W(b|x)^{1/(1+\rho)}q_\rho(b)^{\rho/(1+\rho)}\right\} \right]. \label{eq:lem-SP-optimizers-pf10.2}
\end{equation}
Equations \eqref{eq:lem-SP-optimizers-pf10.1} and \eqref{eq:lem-SP-optimizers-pf10.2} imply \eqref{eq:lem-SP-optimizers-pf10}. We continue with the following assertion:
\begin{lemma}
\begin{equation}
\mE_{\mSP}(R,U_{\cX}) = -\rho_R R + \mD(V_{\rho_R}\|W|U_{\cX}) + \rho_R \mI(U_{\cX};V_{\rho_R}), \label{eq:lem-SP-optimizers-pf12}
\end{equation}
and $V_{\rho_R}$ is a minimizer for $\mE_{\mSP}(R,U_{\cX})$.
\label{cla:claim1}
\end{lemma}

\begin{IEEEproof}
First, note that
\begin{equation}
\mE_{\mSP}(R,U_\cX) = \max_{\rho \in \bbR_+}\left\{ -\rho R + \min_{V \in \cP(\cY|\cX)}\left[ \mD(V\|W|U_{\cX}) + \rho \mI(U_{\cX};V)\right]\right\}, \label{eq:lem-SP-optimizers-pf13}
\end{equation}
which is verified in \cite[Ex.~2.5.23]{csiszar-korner81}. By the subdifferential characterization of Lagrange multipliers (e.g., \cite[Theorem~29.1]{rockafellar70}), $\rho_R$ is the unique maximizer in \eqref{eq:lem-SP-optimizers-pf13}, and hence
\begin{equation}
\mE_{\mSP}(R,U_{\cX}) = -\rho_R R + \min_{V \in \cP(\cY|\cX)}\left\{ \mD(V\|W|U_{\cX}) + \rho_R \mI(U_{\cX};V)\right\}. \label{eq:lem-SP-optimizers-pf-star}	
\end{equation}
Now, for any $\rho \in \bbR_+$,
\begin{align}
\mD\left( V_{\rho}\|W|U_{\cX}\right) + \rho \mI(U_{\cX};V_{\rho}) & = -\ln \sum_{y \in \cY}\left( \sum_{x \in \cX}U_{\cX}(x)W(y|x)^{1/(1+\rho)}\right)^{1+\rho} \\
& = \textnormal{E}_{\textnormal{o}}(\rho, U_{\cX}),\label{eq:lem-SP-optimizers-pf15}
\end{align}
which follows from routine computations once we employ \eqref{eq:SP-optimizers-1} on the right side of \eqref{eq:lem-SP-optimizers-pf10} along with the definition of $q_{\rho}$, i.e., \eqref{eq:lem-SP-optimizers-pf8}. Also, for any $\rho \in \bbR_+$,
\begin{equation}
\min_{V \in \cP(\cY|\cX)}\left[ \mD(V\|W|U_{\cX}) + \rho \mI(U_{\cX};V)\right] \geq  \mE_{\mo}(\rho, U_{\cX}), \label{eq:lem-SP-optimizers-pf16}
\end{equation}
which follows from routine convex analysis arguments (e.g., \cite[Ex.~2.5.23]{csiszar-korner81}). Equations~\eqref{eq:lem-SP-optimizers-pf15} and \eqref{eq:lem-SP-optimizers-pf16}, along with the strict convexity of $\mD(\cdot\|W|U_{\cX})$, which is an immediate consequence of the strict convexity of the function $\bbR_+ \ni x \mapsto x\ln x$, imply that $V_{\rho_R}$ is the unique minimizer in \eqref{eq:lem-SP-optimizers-pf-star}, which, in turn, establishes \eqref{eq:lem-SP-optimizers-pf12}. Since $V_{\rho_R}$ is the unique minimizer in \eqref{eq:lem-SP-optimizers-pf-star}, it must also be primal optimal (e.g., \cite[Theorem~28.1]{rockafellar70}), i.e., it must be a minimizer of $\mE_{\mSP}(R,U_{\cX})$.
\end{IEEEproof}

In order to conclude the proof, consider
\begin{equation}
\me_{\mSP}(R,R)\eqdef \inf_{V \in \cP(\cY|\cX) \colon \mD(V\|q_R|U_{\cX}) \leq R} \mD(V\|W|U_{\cX})
\end{equation}
from~\eqref{eq:eSP-nonsingular}.
By noting the fact that $V_{\rho_R}$ is a minimizer of $\mE_{\mSP}(R,U_{\cX})$, which is verified in Lemma~\ref{cla:claim1}, along with \eqref{eq:lem-SP-optimizers-pf11.1}, we have
\begin{equation}
\mI(U_{\cX};V_{\rho_R}) = \mD(V_{\rho_R}\|q_R|U_{\cX}) \leq R,
\end{equation}
which, in turn, implies that
\begin{equation}
\me_{\mSP}(R,R) \leq \mE_{\mSP}(R,U_{\cX}).
\label{eq:lem-SP-optimizers-pf18}
\end{equation}
Further,
\begin{align}
\me_{\mSP}(R,R) & \geq \sup_{\rho \in \bbR_+} \inf_{V \in \cP(\cY|\cX)}\left\{ \mD(V\|W|U_{\cX}) + \rho \left[ \mD(V\|q_R|U_{\cX}) - R \right] \right\} \label{eq:lem-SP-optimizers-pf19.00} \\
& \geq \inf_{V \in \cP(\cY|\cX)}\left\{ \mD(V\|W|U_{\cX}) + \rho_R \left[ \mD(V\|q_R|U_{\cX}) - R \right]\right\} \label{eq:lem-SP-optimizers-pf19.0}\\
& = \mD(V_{\rho_R}\|W|U_{\cX})+\rho_R \left[ \mD(V_{\rho_R}\|q_R|U_{\cX}) - R \right],\label{eq:lem-SP-optimizers-pf19} \\
& = -\rho_R R + \mD(V_{\rho_R}\|W|U_{\cX}) + \rho_R \mI(U_{\cX};V_{\rho_R}) \label{eq:lem-SP-optimizers-pf19.01}\\
& = \mE_{\mSP}(R, U_{\cX}), \label{eq:lem-SP-optimizers-pf19.02}
\end{align}
where \eqref{eq:lem-SP-optimizers-pf19} follows by solving the convex program in \eqref{eq:lem-SP-optimizers-pf19.0}, \eqref{eq:lem-SP-optimizers-pf19.01} follows from \eqref{eq:lem-SP-optimizers-pf11.1}, and \eqref{eq:lem-SP-optimizers-pf19.02} is \eqref{eq:lem-SP-optimizers-pf12}. Hence, \eqref{eq:lem-SP-optimizers-pf18}, \eqref{eq:lem-SP-optimizers-pf19.00} and \eqref{eq:lem-SP-optimizers-pf19.02} imply that
\begin{align}
\mE_{\mSP}(R,U_{\cX}) = \me_{\mSP}(R,R) & = \max_{\rho \in \bbR_+}\min_{V \in \cP(\cY|\cX)} \left\{ \mD(V\|W|U_{\cX}) + \rho \left[ \mD(V\|q_R|U_{\cX})-R\right]\right\} \label{eq:lem-SP-optimizers-pf20.0}\\
& = \max_{\rho \in \bbR_+}\left\{ -\rho R - (1+\rho)\sum_{x \in \cX}U_{\cX}(x)\ln \sum_{y \in \cY}W(y|x)^{1/(1+\rho)}q_R(y)^{\rho/(1+\rho)}\right\}  \label{eq:lem-SP-optimizers-pf20} \\
& \geq -\rho_R R - (1+\rho_R)\sum_{x \in \cX}U_{\cX}(x)\ln \sum_{y \in \cY}W(y|x)^{\frac{1}{1+\rho_R}}q_R(y)^{\frac{\rho_R}{1+\rho_R}} = \mE_{\mSP}(R,U_{\cX}), \label{eq:lem-SP-optimizers-pf21}
\end{align}
where \eqref{eq:lem-SP-optimizers-pf20} follows by solving the convex program in \eqref{eq:lem-SP-optimizers-pf20.0} and the equality in \eqref{eq:lem-SP-optimizers-pf21} follows from \eqref{eq:lem-SP-optimizers-pf10} and \eqref{eq:lem-SP-optimizers-pf12}. Hence, we conclude that $\rho_R$ attains the supremum in \eqref{eq:lem-SP-optimizers-pf7}.\hfill \IEEEQED
\end{itemize}

\section{Proof of Lemma~\ref{lem:EA}}
\label{app-EA}
Let
\begin{equation}
\hat{S}_N \eqdef \sum_{n=1}^N \frac{Z_n}{N}, 	
\end{equation}
and $\mu_N$ (resp. $\tilde{\mu}_N$) denote the law of $\hat{S}_N$ when $Z_n$ are independent with laws $\nu_n$ (resp. $\tilde{\nu}_{n}$). Let
\begin{equation}
W_N \eqdef  \sum_{n=1}^N \frac{T_n}{\sqrt{m_{2,N}}},	
\end{equation}
where $T_n$ and $m_{2,N}$ are defined right before the statement of the lemma. Via routine change of measure arguments (e.g., \cite[p.~111]{dembo-zeitouni98}), one can check that
\begin{align}
\mu_N\left( [c, \infty) \right) & = \me^{-N \Lambda_N^\ast(c)}\int_{0}^\infty \me^{-x \eta \sqrt{m_{2,N}}}\textnormal{d}F_N(x)  \\
& = \me^{-N \Lambda_N^\ast(c)} \int_{0}^\infty \me^{-t}\left[ F_N\left( \tfrac{t}{\psi_N} \right) - F_N(0)\right] \textnormal{d}t, \label{eq:lem-pf1}
\end{align}
where $F_N$ is the distribution of $W_N$ when $Z_n$ are independent with laws $\tilde{\nu}_{n}$, $\psi_N \eqdef \eta \sqrt{m_{2,N}}$ and \eqref{eq:lem-pf1} follows from an application of the integration by parts. To deduce \eqref{eq:EA-lower-bound}, first note that for any $t \in \bbR_+$
\begin{align}
F_N\left( \tfrac{t}{\psi_N} \right) - F_N(0) & \geq \Phi\left( \tfrac{t}{\psi_N} \right) - \Phi(0) - \frac{2m_{3,N}}{m_{2,N}^{3/2}} \label{eq:lem-pf4.0} \\
&  \geq t\frac{\phi(0)}{\psi_N} - t^2\frac{1}{\psi_N^2 2\sqrt{2 \pi \me}} - \frac{2m_{3,N}}{m_{2,N}^{3/2}}, \label{eq:lem-pf4}
\end{align}
where \eqref{eq:lem-pf4.0} follows from the Berry-Esseen theorem (e.g., \cite[Theorem~III.1]{esseen45}), and \eqref{eq:lem-pf4} follows from a power series approximation, coupled with the observation that $\phi^{\prime}(\cdot) \geq -\frac{1}{\sqrt{2 \pi \me}}$ on $\bbR_+$. Using \eqref{eq:lem-pf4}, we deduce that
\begin{align}
\int_{0}^\infty \me^{-t}\left[ F_N\left( \tfrac{t}{\psi_N} \right) - F_N(0)\right] \textnormal{d}t & \geq \int_{a t_N}^\infty \me^{-t}\left[ F_N\left( \tfrac{t}{\psi_N}\right) - F_N(0)\right] \textnormal{d}t  \\
& \geq \int_{a t_N}^\infty \me^{-t}\left[ \frac{t \left(1 - \tfrac{1}{a}\right)}{\eta \sqrt{2 \pi m_{2,N}}} - \frac{t^2}{\psi_N^2 2 \sqrt{2 \pi \me}}\right]\textnormal{d}t. \label{eq:lem-pf5}
\end{align}
By carrying out the integration on the right side of \eqref{eq:lem-pf5} (e.g., \cite[Eq. (221), (222)]{altug12a}), we conclude that \eqref{eq:EA-lower-bound} holds. \hfill \IEEEQED

\section{Proof of Lemma~\ref{lem:lem2}}
\label{app-lem2}
Define $ S_N \eqdef \sum_{n=1}^N Z_n$ and let $F_N$ denote the distribution function of $S_N$. For convenience, let $B_N(r)$ denote the left side of \eqref{eq:lem2-1} and $m_{1,N} \eqdef \sum_{n=1}^N \mE[Z_n]$. We have
\begin{align}
B_N(r) & = \me^{-r} \int_{-\infty}^{r} \me^{z} \textnormal{d}F_N(z)  \\
& = F_N(r) - \int_{-\infty}^{r}\me^{(z - r)}F_N(z)\textnormal{d}z \label{eq:lem2-pf1}\\
& = \int_{0}^{\infty} \me^{-x}\left[ F_N(r) - F_N\left(r-x\right) \right] \textnormal{d}x  \\
& \leq \int_{0}^{\infty}\me^{-x}\left\{ \int_{\frac{r - m_{1,N}}{\sqrt{m_{2,N}}} - \frac{x}{\sqrt{m_{2,N}}}}^{\frac{r - m_{1,N}}{\sqrt{m_{2,N}}}} \frac{\me^{-\frac{a^2}{2}}}{\sqrt{2 \pi}} \textnormal{d}a + c\frac{ m_{3,N}}{m_{2,N}^{3/2}}\right\}\textnormal{d}x  \label{eq:lem2-pf2}\\
& \leq \frac{1}{\sqrt{2 \pi m_{2,N} }} + c\frac{ m_{3,N}}{m_{2,N}^{3/2}} ,
\end{align}
where \eqref{eq:lem2-pf1} follows from integration by parts, \eqref{eq:lem2-pf2} follows from the Berry-Esseen Theorem\footnote{Similar to earlier invocations, we take the constant in Berry-Esseen theorem as $1$ (resp. $1/2$) if the random variables are independent (resp. i.i.d.), although neither choice is the best possible (e.g., \cite{korolev2010}).} and $c=2$ (resp. $c=1$) if the random variables are independent (resp. i.i.d.).\hfill \IEEEQED

\section{Proof of Lemma~\ref{lem:regularity-nonsingular}}
\label{app-regularity-singular}
We begin by recalling the fact that $(\rho_R, q_R)$ is the unique saddle-point of the right side of \eqref{eq:SP-dual}, which is shown in Proposition~\ref{lem:SP-optimizers-new}(iv), and hence we are in a position to invoke the results proven in \cite{altug12a} throughout the proof.

\begin{enumerate}
\item[(i)] This assertion is a direct consequence of \cite[Lemma~3(ii)]{altug12a}.

\item[(ii)] The claim follows from \cite[Theorem~2]{altug12a}. It was
also shown earlier as part of the proof of 
Proposition~\ref{lem:SP-optimizers-new}(iv)
(see \eqref{eq:lem-SP-optimizers-pf20.0}).

\item[(iii)] First, note that given any $r \in (\mD(W_R\|q_R|U_\cX) , R]$,
\begin{align}
\me_{\mSP}(r,R) & = \max_{\rho \in \bbR_+} \min_{V \in \cP(\cY|\cX)} \left\{ \mD(V\|W|U_{\cX}) + \rho \left( \mD(V\|q_R|U_{\cX}) -r \right)\right\} \label{eq:eSP-nonsingular-pf5} \\
& = \max_{\rho \in \bbR_+}\left\{ - \rho r - (1+\rho)\Lambda\left(\tfrac{\rho}{1+\rho}\right)\right\}, \label{eq:eSP-nonsingular-pf6}
\end{align}
where \eqref{eq:eSP-nonsingular-pf5} follows since the convex program $\me_{\mSP}(r,R)$ has zero duality gap, thanks to the fact that Slater's condition (e.g., \cite[Corollary~28.2.1]{rockafellar70}) holds, which is a direct consequence of the first assertion of this lemma, and \eqref{eq:eSP-nonsingular-pf6} follows by solving the convex program on the right side of \eqref{eq:eSP-nonsingular-pf5}.

The proof of the assertion goes by contradiction. Assume that there exists $\lambda_\mo \in [0,1)$ with $\Lambda^{\prime \prime}(\lambda_\mo)=0$. From \eqref{eq:moments-nonsingular-1} and \eqref{eq:moments-nonsingular-2}, this is equivalent to
\begin{equation}
W(y|x_\mo) = q_R(y) \me^{-\Lambda^\prime(\lambda_\mo)}, \, \forall \, y \in \supp(W(\cdot|x_\mo)).
\label{eq:appD-1}
\end{equation}
Further, \eqref{eq:eSP-nonsingular-pf6} and \eqref{eq:appD-1}, along with the definition of $\Lambda(\cdot)$, imply that
\begin{equation}
\me_{\mSP}(R,R) = \max_{\rho \in \bbR_+} -\rho \left[R + \Lambda^\prime(\lambda_\mo)\right].
\label{eq:appD-1.0}
\end{equation}
Since $\me_{\mSP}(R,R) = \mE_{\mSP}(R)$, which is shown in the second assertion of this lemma, \eqref{eq:appD-1.0} implies that either $\mE_{\mSP}(R) = 0$, which contradicts the fact that $\mE_{\mSP}(R) >0$ (e.g., \cite[p.~158]{gallager68}), or $\mE_{\mSP}(R) = \infty$, which contradicts the fact that $R > R_\infty$. Hence, we conclude that for all $\lambda \in [0,1)$, $\Lambda^{\prime \prime}(\lambda) >0$.

\item[(iv)] For notational convenience, let
\begin{equation}
\me_{\mo}(\rho,R) \eqdef -(1+\rho)\Lambda\left(\tfrac{\rho}{1+\rho}\right).	
\end{equation}
Hence, \eqref{eq:eSP-nonsingular-pf6} reads
\begin{equation}
\me_{\mSP}(r,R) = \max_{\rho \in \bbR_+}\left\{ \me_{\mo}(\rho,R) - \rho r \right\}.
\label{eq:appD-2}
\end{equation}
$\me_{\mSP}(\cdot,R)$ is differentiable owing to \cite[Corollary~2]{altug12a}, and hence we conclude that $s_{(\cdot)}$ is well-defined. Since differentiable convex functions of one variable are continuously differentiable, the second assertion follows. To verify the last two assertions, observe that \eqref{eq:appD-2} is the Lagrangian dual of the convex program $\me_{\mSP}(r,R)$, which is established in \eqref{eq:eSP-nonsingular-pf5} and \eqref{eq:eSP-nonsingular-pf6}. Hence, we can use the subdifferential characterization of the Lagrange multipliers (e.g., \cite[Theorem~29.1]{rockafellar70}) to deduce that the set of optimizers in \eqref{eq:appD-2} coincides with the negative of the subdifferential of $\me_{\mSP}(\cdot,R)$ at $r$, i.e., $\rho \in \bbR_+$ maximizes \eqref{eq:appD-2} if and only if
\begin{equation}
\rho \in - \partial \me_{\mSP}(\cdot, R)(r).	
\end{equation}
Since $\me_{\mSP}(\cdot,R)$ is differentiable at $r$, $-\partial \me_{\mSP}(\cdot, R)(r) = \{ s_r \}$ and hence $s_r$ uniquely attains the maximum in \eqref{eq:appD-2}. Further, since $\me_{\mSP}(r,R) \geq \me_{\mSP}(R,R) = \mE_{\mSP}(R) >0$, we have $s_r \in \bbR^+$.

Moreover, via direct differentiation, one can verify that
\begin{align}
\frac{\partial^2}{\partial \rho^2}\left[ -\rho r + \me_{\mo}(\rho,R)\right] & = \frac{\partial^2 \me_{\mo}(\rho, R)}{\partial \rho^2} \label{eq:appD-3.0}\\
& = -\frac{\Lambda^{\prime \prime}\left( \tfrac{\rho}{1+\rho}\right)}{(1+\rho)^3} \label{eq:appD-3.1}\\
& < 0, \label{eq:appD-3}
\end{align}
where \eqref{eq:appD-3} follows from the third assertion of this lemma. As a direct consequence of \eqref{eq:appD-3}, we conclude that $s_r$ is the unique positive real number satisfying
\begin{equation}
r = \left. \frac{\partial \me_{\mo}(\rho, R)}{\partial \rho} \right|_{\rho = s_r}.	
\end{equation}
This observation, coupled with \eqref{eq:appD-3} and the inverse function theorem, further implies that $s_r$ is strictly decreasing in $r$.

\item[(v)] Since $\Lambda(\cdot)$ is a convex function (e.g., \cite[Lemma~2.2.5(a)]{dembo-zeitouni98}), $\lambda [\me_{\mSP}(r,R) - r] - \Lambda(\lambda)$ is a concave function of $\lambda$ and hence a sufficient condition for $\lambda_\mo \in \bbR$ to attain $\Lambda^\ast(\me_{\mSP}(r,R) - r)$ is
\begin{equation}
\Lambda^\prime(\lambda_\mo) = \me_{\mSP}(r,R) - r.
\label{eq:appD-4}
\end{equation}
As noted above, $s_r$ is the unique positive real number satisfying $r = \left.\frac{\partial \me_{\mo}(\rho, R)}{\rho}\right|_{\rho =s_r}$, hence, an elementary calculation implies that

\begin{equation}
r = - \Lambda\left(\tfrac{s_r}{1+s_r}\right) - \tfrac{1}{(1+s_r)}\Lambda^\prime\left( \tfrac{s_r}{1+s_r}\right),\label{eq:appD-5}
\end{equation}
and hence
\begin{equation}
\me_{\mSP}(r,R) = \tfrac{s_r}{(1+s_r)}\Lambda^\prime\left(\tfrac{s_r}{1+s_r}\right) - \Lambda\left(\tfrac{s_r}{1+s_r}\right).
\label{eq:appD-6}
\end{equation}
Equations~\eqref{eq:appD-5} and \eqref{eq:appD-6} imply that
\begin{equation}
\Lambda^\prime\left(\tfrac{s_r}{1+s_r} \right) = \me_{\mSP}(r,R) -r. \label{eq:appD-7}
\end{equation}
Equation \eqref{eq:appD-7} ensures that $\frac{s_r}{1+s_r}$ attains $\Lambda^\ast(\me_{\mSP}(r,R)-r)$ and hence
\begin{align}
\Lambda^\ast(\me_{\mSP}(r,R) - r) & = \tfrac{s_r}{(1+s_r)}[\me_{\mSP}(r,R) - r] - \Lambda\left(\tfrac{s_r}{1+s_r}\right) \\
& = \me_{\mSP}(r,R), \label{eq:appD-8}
\end{align}
where \eqref{eq:appD-8} follows by substituting \eqref{eq:appD-7} into \eqref{eq:appD-6}.

Finally, let
\begin{equation}
\eta_r \eqdef \tfrac{s_r}{1+s_r},	
\end{equation}
and note that $\eta_r \in \bbR^+$, since $s_r \in \bbR^+$. Hence, \eqref{eq:appD-7} implies the existence of a real number in $(0,1)$, namely $\eta_r$, with
\begin{equation}
\Lambda^\prime(\eta_r) = \me_{\mSP}(r,R) - r. 	
\end{equation}
To verify the uniqueness, it suffices to note that $\me_{\mSP}(\cdot, R) - (\cdot)$ is strictly decreasing, along with the third assertion of this lemma and the inverse function theorem.
\item[(vi)] From the proof of part (iv) we know that $s_R$ is the unique $\rho$ that achieves the maximum in
\begin{align}
\max_{\rho \ge 0} \left\{\me_{\mo}(\rho,R) - \rho R\right\}
  & = \max_{\rho \ge 0} \left\{ -\rho R - (1+\rho) \Lambda\left(\frac{\rho}{1 + \rho}\right)\right\} \\
  & = \max_{\rho \ge 0} \left\{- \rho R - (1+\rho) \ln 
                     \sum_{y \in \mathcal{Y}} q_R(y)^{\rho/(1+\rho)}
                           W(y|x_\mo)^{1/(1+\rho)}\right\}.
\label{eq:srhoequivalence}
\end{align}
But by Proposition~\ref{lem:SP-optimizers-new}(iv) and the 
symmetry of the channel,
$\rho_R$ achieves the maximum in (\ref{eq:srhoequivalence}).
The conclusion follows.
\hfill \IEEEQED
\end{enumerate}

\section{Proof of Lemma~\ref{lem:exponent-nonsingular}}
\label{app-exponent-nonsingular}
The proof follows from essentially the same arguments given in \cite[Section~III.E]{altug12a}. We provide an outline for completeness.

Since $\Lambda(\cdot)$ is smooth (by~\cite[Ex.~2.2.24]{dembo-zeitouni98})
and strictly convex over $(0,1)$ 
(by Lemma~\ref{lem:regularity-nonsingular}(iii)), 
by~\cite[Corollary~23.5.1]{rockafellar70} and the inverse function theorem 
we have that
$\Lambda^\ast(\cdot)$ is twice differentiable over the domain
$$(-\mD(W\|q_R|U_{\cX}) , \mD(W_R\|W|U_\cX))$$ and
\begin{align}
\Lambda^{\ast \, \prime}(\me_{\mSP}(r,R) -r) & = \eta_r, \label{eq:appE-1.0}\\
\Lambda^{\ast \, \prime \prime}(\me_{\mSP}(r,R) -r) & = \frac{1}{\Lambda^{\prime \prime}(\eta_r)}, \label{eq:appE-1.1}
\end{align}
for any $r \in [\bar{R},R]$. Via calculations similar to the ones leading to \cite[Eq. (92)]{altug12a}, one can verify that
\begin{align}
\Lambda^\ast(\me_{\mSP}(R_N,R)-R_N) & = \Lambda^\ast(\me_{\mSP}(R,R)-R) + \varepsilon_N \eta_R + (\me_{\mSP}(R_N,R) - \me_{\mSP}(R,R))\eta_R \nonumber \\
& \quad  + \frac{\Lambda^{\ast \, \prime \prime}(\bar{x})}{2} \left[  \me_{\mSP}(R_N,R)-R_N-\me_{\mSP}(R,R)+R\right]^2, \label{eq:appE-2}
\end{align}
for some $\bar{x} \in (\me_{\mSP}(R,R)-R, \me_{\mSP}(R_N,R)-R_N)$. Using Lemma~\ref{lem:regularity-nonsingular}(iv) and (v), along with the definition of $\varepsilon_N$, \eqref{eq:appE-2} further implies that
\begin{equation}
\me_{\mSP}(R_N,R) = \me_{\mSP}(R,R) + \varepsilon_N s_R + \varepsilon_N^2 (1+s_R)\frac{\Lambda^{\ast \, \prime \prime}(\bar{x})}{2}\left( 1 + \frac{1}{\varepsilon_N}\left[ \me_{\mSP}(R_N,R) - \me_{\mSP}(R,R)\right]\right)^2. \label{eq:appE-3}
\end{equation}
By using \eqref{eq:appE-1.1}, along with the fact that $\me_{\mSP}(\cdot,R) - (\cdot)$ is a strictly decreasing and continuous function over $[\bar{R},R]$, we deduce that
\begin{equation}
\Lambda^{\ast \, \prime \prime}(\bar{x}) \leq \frac{1}{m_{2,\min}} \in \bbR^+.
\label{eq:appE-4}
\end{equation}
Now Lemma~\ref{lem:regularity-nonsingular}(vi) implies that
\begin{equation}
s_R = \rho_R = |\mE_{\mSP}^\prime(R)|.
\label{eq:appE-5}
\end{equation}
Finally, via a first-order power series approximation, along with Lemma~\ref{lem:regularity-nonsingular}(iv) and (v), one can verify that
\begin{equation}
\left( 1 + \frac{1}{\varepsilon_N}\left[ \me_{\mSP}(R_N,R) - \me_{\mSP}(R,R)\right]\right)^2 \leq (1 + s_{\bar{R}})^2.
\label{eq:appE-6}
\end{equation}
Assembling \eqref{eq:appE-3}--\eqref{eq:appE-6}, along with the fact that $\mE_{\mSP}(R) = \me_{\mSP}(R,R)$, which is shown in Lemma~\ref{lem:regularity-nonsingular}(ii), we conclude that \eqref{eq:exponent-nonsingular} holds.\hfill \IEEEQED

\section{Proof of Lemma~\ref{lem:singular-lem1}}
\label{app-singular-lem1}

Similar to the previous sections, for any $x \in \cX$ and $\lambda \in \bbR$, define
\begin{equation}
M_x(\lambda) \eqdef \sum_{y \in \supp(W(\cdot|x))}W(y|x)^{1-\lambda} q(y)^\lambda.	
\end{equation}
Evidently, $M_x(\cdot) \in \bbR$ for any $x \in \cX$.

Next, we claim that given any $\lambda \in \bbR$, $M_x(\lambda)$ is constant in $x$, whose proof is similar to Lemma~\ref{lem:iid-nonsingular}(i). Specifically, let $\{ \cY_l\}_{l=1}^L$ be a partition of the columns of $W$ mentioned in Definition~\ref{def:symmetric}, whose choice is immaterial in what follows. Since each column is a permutation of every other column for any sub-channel defined by this partition, $q(y)$ is the same for any $y \in \cY_l$. This observation, along with the fact that every row is a permutation of every other row for any sub-channel defined by the aforementioned partition, implies that $M_x(\cdot)$ is the same for all $x \in \cX$.

\begin{itemize}
\item[(i)] By noting the fact that whenever $W(y|x)>0$,
\begin{equation}
\frac{W(y|x)}{q(y)} = \frac{1}{\alpha_y},	
\end{equation}
which is a direct consequence of the fact that $W$ is singular, we deduce that
\begin{equation}
\sum_{\by^N \in \cS(r)} \prod_{n=1}^N W(y_n | \psi_n(\by^{n-1})) = \sum_{\by^N \in \cY^N}\prod_{n=1}^N W(y_n | \psi_n(\by^{n-1})) \b1\left\{ \frac{1}{N}\sum_{n=1}^N \ln \frac{W(y_n| \psi_n(\by^{n-1}))}{q(y_n)} \leq r \right\}, \label{eq:appF-2.0}
\end{equation}
where $\psi_1(\by^0)$ denotes $\psi_1$. Next, similar to the proof of Lemma~\ref{lem:iid-nonsingular}(ii), one can check that for any $\lambda \in \bbR$,
\begin{equation}
\sum_{\by^N \in \cY^N} \prod_{n=1}^N W(y_n | \psi_n(\by^{n-1})) e^{\lambda \ln \prod_{n=1}^N \frac{q(y_n)}{W(y_n | \psi_n(\by^{n-1}))}}= M_{x_\mo}(\lambda)^N. \label{eq:appF-2.1}
\end{equation}
Using the uniqueness theorem for the moment generating function (e.g., \cite[Ex.~26.7]{billingsley95}), \eqref{eq:appF-2.0} and \eqref{eq:appF-2.1} suffice to conclude the assertion.

\item[(ii)] Define
\begin{align}
\Lambda(\lambda) & \eqdef \ln \mE_{W(\cdot|x_\mo)}\left[ e^{\lambda \ln \frac{q(Y)}{W(Y|x_\mo)}}\right] \\
 & = \ln \sum_{y \in \supp(W(\cdot|x_\mo))} W(y|x_\mo)^{1-\lambda} q(y)^\lambda.	
\end{align}
The singularity of $W$, along with \eqref{eq:singular-conv1}, implies that
\begin{equation}
\Lambda(\lambda) = \ln \sum_{y \in \supp(W(\cdot|x_\mo))} \xi_y \alpha_y^\lambda.
\label{eq:appF-3}
\end{equation}
Observe that for any $\lambda \in \bbR_+$,
\begin{align}
\Lambda(\lambda) & = \ln \sum_{y \in \cY}\xi_y \alpha_y^{1+\lambda} \label{eq:appF-4.0}\\
& = -\mE_{\mo}(\lambda, U_{\cX}), \label{eq:appF-4}
\end{align}
where $\mE_{\mo}(\cdot,\cdot)$ is defined in \eqref{eq:Eo}, \eqref{eq:appF-4.0} follows from Proposition~\ref{lem:SP-optimizers-new}(ii) and \eqref{eq:appF-4} follows from an elementary calculation by noticing the singularity of the channel. Note that \eqref{eq:appF-4} enables us to relate
\begin{equation}
\Lambda^\ast(-R) \eqdef  \sup_{\lambda \in \bbR}\left\{ -\lambda R - \Lambda(\lambda) \right\}
\end{equation}
to $\mE_{\mSP}(R)$, and hence is the crucial step of the proof. Moreover, it relies on the singularity of the channel.

Continuing with the proof, one can check that
\begin{align}
\Lambda^\prime(\lambda) & = \sum_{y \in \supp(W(\cdot|x_{\mo}))} \frac{\xi_y \alpha_y^{\lambda}}{\sum_{b \in  \supp(W(\cdot|x_{\mo}))} \delta_b \alpha_b^{\lambda}}\ln \alpha_y, \label{eq:appF-5.0}\\
\Lambda^{\prime \prime}(\lambda) & = \sum_{y \in \supp(W(\cdot|x_{\mo}))} \frac{\xi_y \alpha_y^{\lambda}}{\sum_{b \in \supp(W(\cdot|x_{\mo}))} \delta_b \alpha_b^{\lambda}} \left(  \ln \alpha_y - \Lambda^\prime(\lambda) \right)^2 \label{eq:appF-5.1-0}\\
& \geq 0, \label{eq:appF-5.1-1}	
\end{align}
for any $\lambda \in \bbR_+$. Further, define
\begin{equation}
m_{3}(\lambda) \eqdef  \sum_{y \in \supp(W(\cdot|x_{\mo}))} \frac{\xi_y \alpha_y^{\lambda}}{\sum_{b \in \supp(W(\cdot|x_{\mo}))} \delta_b \alpha_b^{\lambda}}\left| \ln \alpha_y - \Lambda^{\prime}(\lambda) \right|^3.
\label{eq:appF-7}
\end{equation}
Evidently, $\Lambda^\prime(\cdot), \Lambda^{\prime \prime}(\cdot)$ and $m_3(\cdot)$ are bounded and continuous over $\bbR_+$. Next, we prove that
\begin{equation}
\forall \, \lambda \in \bbR_+, \, \, \Lambda^{\prime \prime}(\lambda) > 0.
\label{eq:appF-6}
\end{equation}
In order to see \eqref{eq:appF-6}, first note that
\begin{equation}
\Lambda^{\prime \prime}(\lambda) \geq 0, \, \forall \, \lambda \in \bbR_+,	
\end{equation}
due to \eqref{eq:appF-5.1-1}. Assume there exists $\lambda_\mo \in \bbR_+$ with $\Lambda^{\prime \prime}(\lambda_\mo)=0$. This, however, implies that $R_{\mcr}=C(W)$, owing to \eqref{eq:appF-4}, \cite[Theorem~5.6.3]{gallager68}, Remark~\ref{rem:misc}(i) and the fact that $U_\cX$ is a capacity achieving input distribution for $W$, which yields a contradiction.

For any $r \in (R_\infty, R]$, let
\begin{equation}
\rho_r \eqdef -\left.\frac{\partial \mE_{\mSP}(a,U_{\cX})}{\partial a}\right|_{a = r},	
\end{equation}
which is a well-defined mapping owing to \cite[Proposition~3]{altug12a}. Further, observe that for any $r \in (R_\infty,R]$,
\begin{equation}
-r = \Lambda^\prime(\rho_r),
\label{eq:appF-0}
\end{equation}
which is evident in light of
\begin{align}
r & = \left.\frac{\partial \mE_{\mo}(\rho, U_{\cX})}{\partial \rho}\right|_{\rho = \rho_r} \label{eq:appF-0.0}\\
& = - \Lambda^{\prime}(\rho_r), \label{eq:appF-0.00}
\end{align}
where \eqref{eq:appF-0.0} follows by recalling the fact that $\rho_r$ attains $\tilde{\mE}_{\mSP}(r,U_{\cX})$, which is shown in Proposition~\ref{lem:SP-optimizers-new}(iii), and \eqref{eq:appF-0.00} follows from \eqref{eq:appF-4}. Moreover, since $\rho_r$ attains $\tilde{\mE}_{\mSP}(r,U_{\cX})$ and for any $r \in (R_\infty,R]$,
\begin{equation}
\tilde{\mE}_{\mSP}(r, U_{\cX}) \geq \tilde{\mE}_{\mSP}(R, U_{\cX}) = \tilde{\mE}_{\mSP}(R) >0,	
\end{equation}
we deduce that $\rho_r \in \bbR^+$. Further, \eqref{eq:appF-6}, \eqref{eq:appF-0} and the inverse function theorem ensure that $\rho_{(\cdot)}$ is strictly decreasing over $(R_\infty,R]$.

To conclude the proof, we fix some $a>1$ and define
\begin{align}
t_{\max} & \eqdef a 2\sqrt{2 \pi} \rho_{\bar{R}} \max_{\lambda \in [0, \rho_{\bar{R}}]} \frac{m_3(\lambda)}{\Lambda^{\prime \prime}(\lambda)} , \\
m_{2, \min} & \eqdef \min_{\lambda \in [0, \rho_{\bar{R}}]} \Lambda^{\prime \prime}(\lambda), \\
m_{2, \max} & \eqdef \max_{\lambda \in [0, \rho_{\bar{R}}]} \Lambda^{\prime \prime}(\lambda),
\end{align}
where $\bar{R}=\tfrac{R+R_\infty}{2}$, as defined before. Clearly, all of the above are well-defined and positive quantities. For convenience, let
\begin{equation}
\frac{\me^{-t_{\max}} \left(1 - \tfrac{1}{a} \right)}{\rho_{\bar{R}}  2 \sqrt{ 2 \pi m_{2, \max}}} =\mathrel{\mathop:} k_\mo  \in \bbR^+.
\end{equation}
Let $N \in \bbZ^+$ be sufficiently large such that
\begin{align}
R_N & \geq \bar{R}, \label{eq:appF-8.0} \\
\frac{1+(1+t_{\max})^2}{\rho_{\bar{R}}\left( 1 - \frac{1}{a} \right) 2 \sqrt{\me N  m_{2, \min}}} & \leq 1/2, \label{eq:appF-8.00}
\end{align}
and note that
\begin{align}
W\left\{ \cS(R_N) | \bx^N_{\mo}\right\} & \geq k_\mo \left(1+ a 2 \sqrt{2 \pi} \rho_{R_N} \frac{m_3(\rho_{R_N})}{\Lambda^{\prime \prime}(\rho_{R_N})}\right)\frac{1}{\sqrt{N}}\me^{-N \Lambda^\ast(-R_N)} \label{eq:appF-bound0} \\
& \geq \frac{k_\mo}{\sqrt{N}}\me^{-N \Lambda^\ast(-R_N)},\label{eq:appF-bound1}
\end{align}
where \eqref{eq:appF-bound0} follows from Lemma~\ref{lem:EA}, which is applicable thanks to \eqref{eq:appF-6} and \eqref{eq:appF-0}, along with \eqref{eq:appF-8.0} and \eqref{eq:appF-8.00}. Since $\rho_{(\cdot)} \in \bbR^+$ is strictly decreasing and $\Lambda(\cdot)$ is convex, \eqref{eq:appF-0} implies that
\begin{align}
\Lambda^\ast(-R_N) & = \max_{0 \leq \lambda  \leq \rho_{\bar{R}}}\left\{ - \lambda \left( R - \frac{k}{N} \right)- \Lambda(\lambda) \right\} \\
&  \leq \frac{k \rho_{\bar{R}}}{N} + \max_{0 \leq \lambda \leq \rho_{\bar{R}}}\left\{ - \lambda  R - \Lambda(\lambda) \right\} \\
&  \leq \frac{k \rho_{\bar{R}}}{N} + \sup_{\lambda \in \bbR_+ }\left\{ - \lambda  R - \Lambda(\lambda) \right\}\\
& = \frac{k \rho_{\bar{R}}}{N} +  \sup_{\lambda \in \bbR_+ }\left\{ - \lambda  R + \mE_{\mo}(\lambda, U_{\cX}) \right\} \label{eq:appF-bound3.0} \\
&  = \frac{k \rho_{\bar{R}}}{N} + \mE_{\mSP}(R),
\label{eq:appF-bound3}
\end{align}
where \eqref{eq:appF-bound3.0} follows from \eqref{eq:appF-4} and \eqref{eq:appF-bound3} follows from Proposition~\ref{lem:SP-optimizers-new}(i). By substituting \eqref{eq:appF-bound3} into \eqref{eq:appF-bound1}, we deduce the assertion.\hfill \IEEEQED
\end{itemize}

\section*{acknowledgment}
The first author thanks Emre Telatar and Paul Cuff for their hospitality while portions of this work were being completed during his visits to \'EPFL and Princeton University. The authors thank Sergio Verd\'u for raising the question of whether the proof methodology of Theorem~\ref{thrm:singular} can be used in the fixed-error probability regime.
This research was supported by the National Science Foundation under grants CCF-1218578 and CCF-1513858.

\end{document}